\documentclass[a4paper,pra,reqno,superscriptaddress,twocolumn,
showpacs,footinbib]{revtex4-2}
\usepackage[utf8]{inputenc}
\usepackage{amsmath}
\usepackage{amsfonts}
\usepackage{dsfont} 
\usepackage{amssymb}
\usepackage{makeidx}
\usepackage{graphicx}
\usepackage{color}
\usepackage{mathptmx}
\usepackage{changes,enumitem,todonotes}
\usepackage{soul}
\usepackage[mathscr]{euscript}
\usepackage{tikz}
\usepackage[normalem]{ulem}
\usepackage[utf8]{inputenc}
\usepackage{amsmath}
\usepackage{amsfonts}
\usepackage{amssymb}
\usepackage{subcaption} 
\usepackage{hyperref}
\hypersetup{colorlinks=true,linkcolor=blue,urlcolor=blue,citecolor=blue} 
\usepackage{mathtools} 

\begin{document}
\title{Quantum light-matter interactions in structured waveguides}
\author{Rupak Bag$^1$ and Dibyendu Roy}
\affiliation{ Raman Research Institute, Bengaluru 560080, India}

\begin{abstract}
We explore special features of quantum light-matter interactions inside structured waveguides due to their finite bandwidth, band edges, and non-trivial topological properties. We model the waveguides as either a tight-binding (TB) chain or a Su-Schrieffer-Heeger (SSH) chain. For unstructured waveguides with infinite bandwidth, the transmission and reflection amplitude of a side-coupled two-level emitter (2LE) are the same as the reflection and transmission amplitude of a direct-coupled 2LE. We show that this analogy breaks down for structured waveguides with finite bandwidth due to the appearance of Lamb shift only for the direct-coupled 2LE. We further predict a robust light-emitter coupling at zero collective decay width of a single giant 2LE (with two couplings at different points) near the band edges of the structured waveguides where topological features can be beneficial. Finally, we study single-photon dynamics in a heterojunction of a long TB and short SSH waveguide connected to a 2LE at the SSH end. We show the propagation of a photon from the excited emitter to the TB waveguide only when the SSH waveguide is in the topological phase. Thus, the heterojunction acts as a quantum switch or conditional propagation channel.      
\end{abstract}
\maketitle

\section{Introduction}
Waveguide quantum electrodynamics (WQED)\cite{RevModPhys.89.021001, RevModPhys.90.031002,gu2017microwave,RevModPhys.95.015002} is a descendant of cavity QED, with no optical confinement along the propagation direction of light. Strong light-matter interactions are generated in the WQED systems by employing tight confinement of the light fields to deeply subwavelength sizes in the transverse dimensions using open waveguides and/or large effective dipole moment of the emitter(s). While a linear energy-momentum dispersion with an infinite bandwidth for propagating photons inside the continuum waveguides has been extensively investigated for physical reasons and convenience \cite{RevModPhys.89.021001,RevModPhys.95.015002}, some previous studies have also explored the nonlinear dispersion with a finite bandwidth for structured waveguides \cite{PhysRevLett.101.100501, PhysRevLett.104.023602, PhysRevA.83.043823, PhysRevLett.115.063601, Schmidt13, Goban14, HoodKimble2016, LiuHouck2017, Fitzpatrick2017, mirhosseini2018superconducting, PhysRevLett.129.093602, vrajitoarea2022ultrastrong}. Both single- and multi-photon transport were investigated inside structured waveguides with a tight-binding (TB) chain dispersion, respectively, for quantum switching \cite{PhysRevLett.101.100501} and the effect of finite bandwidth on generated photon-photon interaction via matter \cite{PhysRevLett.104.023602, PhysRevA.83.043823}. Further, recent studies have explored giant emitters coupled to structured waveguides \cite{zhao2020single,PhysRevA.107.013710}. Nevertheless, this paper shows that many exciting features of light-matter interactions in structured waveguides have yet to be found earlier. Notably, we demonstrate here that a well-known analogy \cite{PhysRevA.79.023837} between the transport coefficients through a two-level emitter (2LE) side coupled and direct coupled to continuum waveguides breaks down for structured waveguides. It is due to the appearance of finite Lamb shift \cite{mirhosseini2018superconducting, PhysRev.72.241, PhysRev.72.339} in a 2LE direct coupled to structured waveguides of finite bandwidth. The Lamb shift is absent for such a configuration of an infinite bandwidth. There is no Lamb shift for the 2LE side coupled to structured or continuum waveguides with finite or infinite bandwidth, respectively.

Topological features of structured waveguides can further influence the light-matter interactions \cite{ozawa2019topological}. 
\textcite{bello2019unconventional} have investigated interactions of single or multiple emitters with light confined inside a structured waveguide modeled as a Su-Schrieffer-Heeger (SSH) lattice with non-trivial band topology. They discovered interesting chiral (spatially asymmetric) emission of an excited side-coupled 2LE to the SSH waveguide when the transition energy of the 2LE lies in the band gap of the waveguide \cite{bello2019unconventional, PhysRevA.104.053522}. The chirality of emitted photon depends on the position of the emitter's coupling to the particular type of sublattice of the SSH waveguide. These predictions were later experimentally verified by \textcite{PhysRevX.11.011015}.
\begin{figure*}
\includegraphics[height=5.7cm,width=16.0cm]{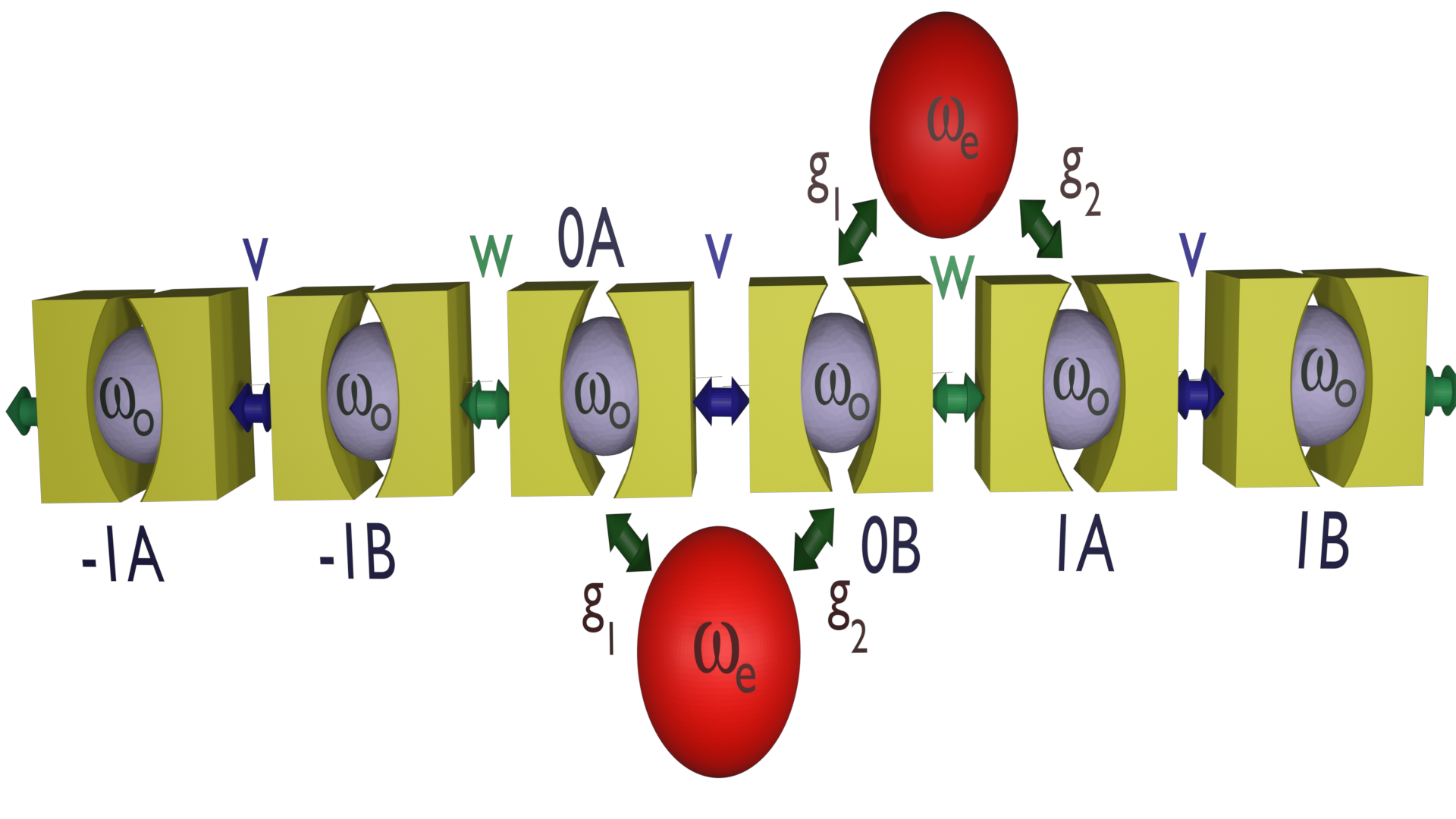}
\caption{Two configurations of a giant 2LE of transition frequency $\omega_e$ side coupled to two neighboring sites of an SSH waveguide modeled by the resonators of frequency $\omega_0$ connected via hopping amplitudes $v,w$. The configurations $A$-$B$ and $B$-$A$ are defined by the connection sites ($0A$,$0B$) and ($0B$,$1A$), respectively, with coupling amplitudes $g_1$ and $g_2$. }\label{G2LESSH}
\end{figure*}

For WQED with superconducting circuits, an emitter can be capacitively connected to multiple discrete points of the superconducting transmission line waveguide. While each coupling can be treated within the dipole and rotating-wave approximation (RWA), the resultant coupling between light in the transmission line and emitter breaks the dipole approximation to produce an effective large or giant emitter \cite{PhysRevA.90.013837} when the distance between the discrete coupling points is comparable to the resonant light wavelength. Such giant emitter displays many remarkable features, such as frequency-dependent energy (Lamb) shifts and decay widths (non-Markovian behavior) and waveguide-mediated decoherence-free interactions between emitters \cite{PhysRevLett.120.140404,kannan2020waveguide}. Interaction of a giant 2LE with an SSH waveguide has been recently explored to understand the features of chiral emission from excited giant emitter within the band gaps of the waveguide, and the role of multiple coupling to different sublattice sites \cite{PhysRevA.106.033522}. Further, \textcite{bello2019unconventional} calculated single-photon transmission lineshape through a giant 2LE coupled to both sublattices of the same unit cell of the SSH waveguide for studying the nature of transmission dip at different Lamb shifts for different topological phases of the waveguide. Interestingly, the Lamb shift for such a giant emitter configuration (two couplings in the same unit cell) is independent of the incident light's frequency (momentum) and the relative phase factor between two sublattices in the Bloch states of the SSH waveguide. The latter phase factor determines the topological character of the waveguide. The phase can instead be extracted from the decay width of the giant 2LE in such a configuration. 
 
 Here we carefully study the decay width, Lamb shift, and transport coefficients through a giant 2LE for two different configurations of the giant emitter. These configurations are defined by two waveguide-emitter couplings to different sublattice sites belonging to the same or different unit cell(s) of the SSH waveguide as depicted in Fig.~\ref{G2LESSH}. We use the band edges of the structured waveguides \cite{LiuHouck2017, HoodKimble2016} to predict a robust light-emitter coupling (i.e., a significant Lamb shift) at zero collective decay width of a giant 2LE, indicating a decoherence-free strong coupling \cite{PhysRevLett.120.140404, mirhosseini2018superconducting}. Such strong coupling is possible for any (odd) number of unit cell separations between two couplings of the emitter with an SSH (TB) waveguide. We also show a constraint ({\it no-go theorem}) on the collective decay rate of the giant 2LE in one of the two topological phases of the SSH chains for these configurations.
      
 Finally, we explore single-photon dynamics in a heterojunction made of a finite SSH waveguide connected to a long TB waveguide. Applying the edge modes of a finite SSH waveguide, \textcite{PhysRevX.11.011015} demonstrated the quantum state transfer between distant qubits attached to the ends of the SSH waveguide. We apply a similar concept to show the conditional propagation of a photon to the TB waveguide from an excited emitter connected to the other end of the SSH waveguide. The photon travels to the TB waveguide in finite time only when the SSH waveguide is in the topological phase. We study the scattering properties in these waveguide QED models employing the Lippmann-Schwinger (LS) scattering theory \cite{PhysRevA.83.043823} and the direct simulation of time-dependent Schr{\"o}dinger equation. The rest of the article is divided into four sections for the main results and discussions and four appendices for details of the non-trivial calculations. 

\section{Decay width and Lamb shift of a giant emitter}\label{decay}
We consider two lattice models for the structured waveguide, namely, the TB and the SSH model. While both these models give a nonlinear energy-momentum  dispersion and a finite bandwidth for propagating photons in the waveguide, the SSH model also displays nontrivial topological features and a band gap. We extract the decay rate and Lamb shift of a giant 2LE coupled to these structured waveguides by finding the light scattering properties in these systems applying the LS formalism. Let us denote $\hat{H}_w, \hat{H}_e$ and $\hat{H}_i$ for the Hamiltonian of the light fields inside waveguide, the 2LE and the interaction between the light fields and 2LE, respectively. In the LS formalism, an eigenstate $|\psi\rangle$ of the full Hamiltonian $\hat{H}=\hat{H}_w+\hat{H}_e+\hat{H}_i$ with energy $E$, is related to eigenstate $| \tilde \phi\rangle$ of the free Hamiltonian  $\hat{H}_o=\hat{H}_w+\hat{H}_e$ with the same energy $E$ via the following relation:
\begin{align}
    |\psi\rangle=|\tilde \phi\rangle+G_o^{R}(E)\hat{H}_i|\psi\rangle, G_o^{R}(E)=\lim_{\epsilon\rightarrow0}\;\frac{\mathds 1}{E-\hat{H}_o+ i\epsilon}, \label{LippmannRetardG}
 \end{align}
where $G_o^{R}(E)$ is the retarded Green's function of the free system.

\subsection{TB waveguide}\label{GiantTB}
The Hamiltonian for bosonic light fields in the TB lattice after setting $\hbar=1$ is 
\begin{align}\label{eq:TB hamiltonian}
\hat{H}_{TB} = \sum_{x=-\infty }^{ \infty} J\big( c_{x}^{\dagger} c_{x+1}+ c_{x+1}^{\dagger}c_{x} \big)+\omega_o c_{x}^{\dagger} c_{x},
\end{align}
where, $c_{x}^{\dagger}$ ($c_{x}$) is the photon creation (annihilation) operator at $x^{\text{th}}$ site, $\omega_o$ is the onsite energy, and $J$ is the hopping amplitude for a photon between nearest neighbor sites. Each site of the lattice can be conceived as a resonator, and the hoppings are generated due to evanescent-field coupling or evanescent Bloch waves \cite{yariv1999coupled}. The energy-momentum dispersion relation of $\hat{H}_{TB}$ is  $\omega_k=\omega_o+2J\cos{k}$ with wave vector $k\in [-\pi,\pi)$. The sinusoidal energy-momentum dispersion can be implemented in coupled-resonator optical waveguides \cite{yariv1999coupled} or photonic crystal structures \cite{PhysRevLett.115.063601, HoodKimble2016}. We take a 2LE being side coupled at two distinct sites, $x=0,\Delta x$ (with $\Delta x>0)$ of the TB waveguide with amplitude $g_1$, $g_2$, respectively. The Hamiltonian of the 2LE with the frequency $\omega_g$ and $\omega_e$ for the ground ($|g\rangle$) and excited ($|e\rangle$) state is,
\begin{align}\label{TLS Hamiltonian}
    \hat{H}_e=\omega_g\sigma \sigma^{\dagger} +\omega_e\sigma^{\dagger} \sigma,
\end{align}
where $\sigma=|g\rangle\langle e|$ and $\sigma^{\dagger}=|e\rangle\langle g|$, and we set $\omega_g=0$. The interaction Hamiltonian is written assuming the RWA: 
\begin{align}
     \hat{H}_i =(g_1 c_{0}^{\dagger}+g_2 c_{\Delta x}^{\dagger})\sigma + \sigma^{\dagger}(g_1 c_{0}+g_2 c_{\Delta x}).\label{intGTB}
\end{align}
We then consider a single-photon input state $\phi(x)$ in the waveguide from the left of the emitter propagating towards right, $\phi(x)\equiv \langle x|\phi \rangle=e^{ikx}/\sqrt{2\pi}$ and the emitter in its ground state. 

We only take negative values of $k$ since the group velocity $v_g(k)=\frac{\partial\omega_k}{\partial k}=-2J\sin{k}>0$ for $k<0$. We apply the LS formalism to derive the single-photon transmission ($T_k$) and reflection ($R_k$) coefficient as following:
\begin{align}
    T_k=1-R_k,\;R_k=\frac{(\Gamma_k/2)^2}{(\omega_k-\omega_e-\Delta_k)^2+(\Gamma_k/2)^2}.\label{reflection coeff TB}
\end{align}
The giant 2LE acts a perfect mirror $(R_k=1)$ for a resonant frequency of the incident light $\omega_k=\omega_e+\Delta_k$, where the wavevector-dependent (for $\Delta x >1$) Lamb shift is 
\begin{align}
    \Delta_k[g_1,g_2]=\frac{ 2}{v_g(k)} g_1 g_2 \sin{k \Delta x}. \label{TBeshift}
\end{align}
The width of the reflection lineshape near the resonant frequency is determined by $\Gamma_k$ (the decay rate), is also wavevector dependent,
\begin{align}
    \Gamma_k[g_1,g_2]=\frac{2}{v_{g}(k)}\Big(g_1^2+g_2^2+2g_1 g_2 \cos{(k \Delta x)}\Big).\label{TBdecay}
\end{align}

Both the Lamb shift $\Delta_k$ and the decay width $\Gamma_k$ can be probed experimentally. The decay width  gives a measure of the quantum interference effect in the giant emitter configuration arising due to interference between two paths of light propagation for $x\in [0,\Delta x]$: one through the emitter and another through the waveguide. The argument of the cosine term in Eq.~\ref{TBdecay} contains the phase picked up by light traveling an extra distance $\Delta x$. Nevertheless, these quantum interference effects can be suppressed to zero by properly fixing the wavevector of the incident light within the TB bandwidth associated with a given giant emitter configuration, i.e., there is a valid  solution of $\cos(k\,\Delta x)=0$ in the physical domain of $k\in [-\pi,0]$ for any $\Delta x$ (including $\Delta x=1$). In the absence of interference between two paths for $k\Delta x=-\pi/2$, we thus have $\Gamma_k[g_1,g_2]=\Gamma_k[g_1,0]+\Gamma_k[0,g_2]$, i.e., the total decay width of the giant 2LE is a sum of the individual decay width. We later discuss a qualitative change in the discussed behavior of decay width of a giant emitter for a topological waveguide. 
\begin{figure}
\includegraphics[width=0.47\textwidth]{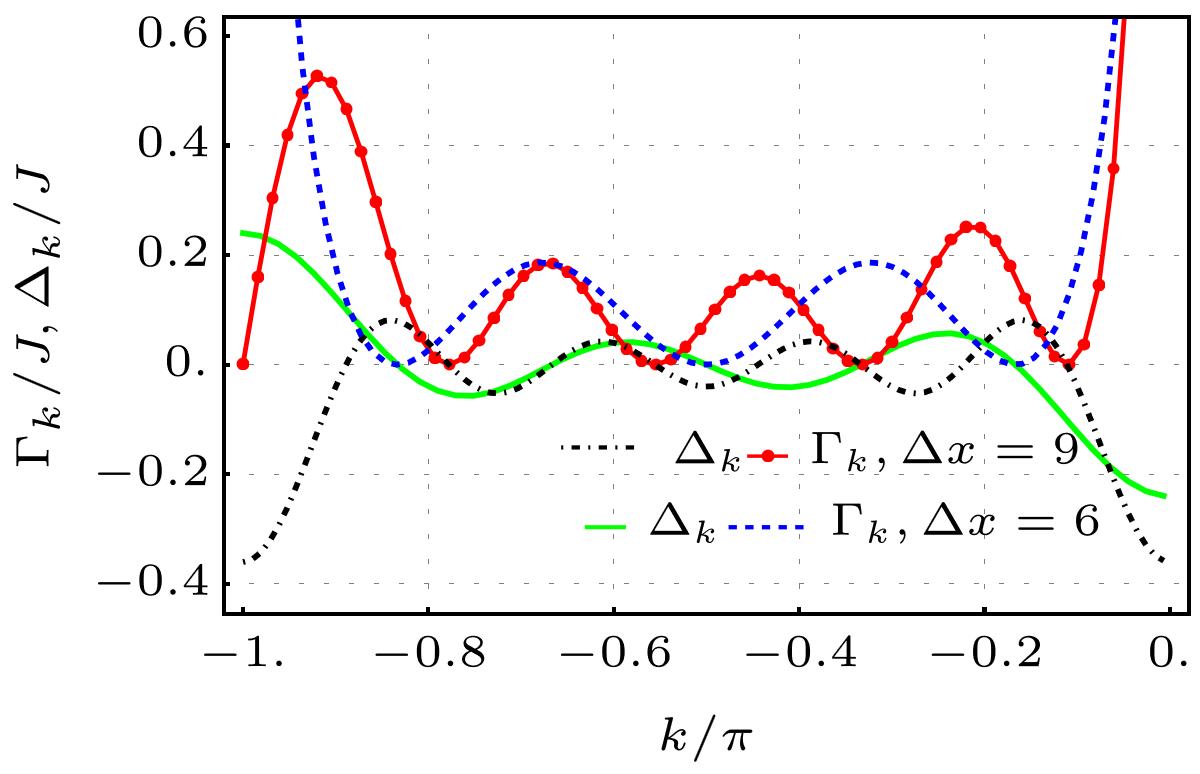}
\caption{Wave vector $(k)$ dependence of decay rate $\Gamma_k$ and Lamb shift $\Delta_k$ of a giant 2LE side coupled to a TB waveguide at two distinct sites with an even (blue dotted and green full lines) and an odd (red solid-dotted and black dash-dotted lines) unit cell separation $\Delta x$. Parameters are $g_1/J=g_2/J=0.2$.} \label{TBgiant}
\end{figure}

For such 1D structured waveguides, the photon group velocity $v_g(k)$ approaches zero near the band edges for $k=0,\pm\pi$ (even though optical phase velocity $v_p(k)=\omega_k/k$ can be non zero or tend towards infinity at the band edges), which leads to an exceedingly long optical path length in the waveguide. This in turn enhances the individual decay widths, $2g_1^2/v_{g}(k), 2g_2^2/v_{g}(k)$, near the band edges since photons spend longer time around the emitter. In the recent years, the giant emitters have been explored for decoherence-free interaction between them \cite{PhysRevLett.120.140404}. Here, we show that a large Lamb shift of a giant 2LE with zero total decay width ($\Gamma_k[g_1,g_2]=0$) can be generated near the band edges of the TB waveguide \cite{LiuHouck2017, HoodKimble2016, mirhosseini2018superconducting}. A large $\Delta_k$ also indicates a strong light-emitter coupling. We find from Eqs.~\ref{TBeshift}, \ref{TBdecay} that $\Gamma_k[g_1,g_2]=0$ implies $\cos{(k\,\Delta x)}=-1$, i.e., $k_n=-(2n+1)\pi/\Delta x$ where $n=0,1,2,\dots$ for $g_1=g_2$. Interestingly, $\Delta_k[g_1,g_2]$ also vanishes at these $k_n$ excluding $k_n$ around $-\pi$ for odd integer values of $\Delta x$. Therefore, for an odd integer unit cells separation between the two couplings  of a giant 2LE with a TB waveguide, we have $\Gamma_{k \to -\pi}[g_1,g_1]=0$ and $\Delta_{k \to -\pi}[g_1,g_1]=-g_1^2\Delta x/J$ as $\lim_{k\to -\pi}\sin{k \Delta x}/\sin{k}=(-1)^{\Delta x +1}\Delta x$. We display the features of $\Gamma_k[g_1,g_2]$ and $\Delta_k[g_1,g_2]$ for odd and even $\Delta x$ in Fig.~\ref{TBgiant}. We show next that such a strong light-emitter coupling (i.e., a large $\Delta_k$) at zero collective decay width of a giant 2LE (i.e., a decoherence-free strong coupling) is possible for both an even and an odd integer unit cells separation between two couplings of the emitter with an SSH waveguide.

\subsection{SSH waveguide}\label{GiantSSH}
The Hamiltonian for photon fields in the SSH lattice is:
\begin{align}
   \hat{H}_{SSH} =& \sum_{x=-\infty}^{\infty} \omega_o\big(a_{x}^{\dagger} a_{x} + b_{x}^{\dagger} b_{x}\big)+ v\big(a_{x}^{\dagger} b_{x}+ b_{x}^{\dagger} a_{x} \big)\nonumber \\ &+w \big(a_{x+1}^{\dagger} b_{x}+ b_{x}^{\dagger} a_{x+1}\big). \label{SSH Hamiltonian}
\end{align}
Here, $a_{x}^{\dagger}$ ($b_{x}^{\dagger}$) is photon creation operator at the sublattice site $A$ ($B$) of $x^{\text{th}}$ unit cell. The hopping amplitude inside (between) the unit cell(s) is $v \equiv J(1-\delta)$ ($w\equiv J(1+\delta) $), and the onside frequency for both sublattices is $\omega_o$. The SSH lattice has two energy bands separated by a bulk-gap for $v\neq w$. The bulk dispersion and eigenvectors for these bands are, respectively,  
\begin{align}
    \omega_k^{\pm}&=\omega_o \pm \sqrt{v^2+w^2+2vw\, \cos{k}} \equiv \omega_o \pm f_k,\label{SSH dispersion} \\
    |\phi_k^{\pm}\rangle&=\frac{1}{\sqrt{4\pi}}\sum_{x=-\infty}^{\infty}e^{ik\,x}\Big(\pm e^{-i\theta_k}|x,A\rangle+|x,B\rangle\Big),\label{SSH Eigen states}
\end{align} 
where, $k\in [-\pi,\pi)$, $|x,A\rangle \equiv |x\rangle \otimes |A\rangle=a_x^{\dagger}|\varphi\rangle, |x,B\rangle\equiv |x\rangle \otimes |B\rangle=b_x^{\dagger}|\varphi\rangle$. The superscript ($\pm$) denotes upper/lower band, and $|\varphi\rangle$ indicates the vacuum mode of the photon fields. The $k$-dependent relative phase factor $\theta_k$ between two sublattices in the Bloch states captures the essence of bulk-topology of the SSH lattice. 
\begin{align}
 \theta_k =\text{Arg}\Big[v+we^{ik}\Big].\label{topology dependent phase}
\end{align}
The Zak phase $\gamma_{\pm}$, which is a bulk topological invariant of the SSH model \cite{asboth2016short}, is related to $\theta_k$ as
\begin{align}
    \gamma_{\pm}=i\int_{-\pi}^{\pi} dk\langle u_k^{\pm}|\partial_k|u_k^{\pm}\rangle=\frac{1}{2}\big(\theta_{k=\pi}-\theta_{k=-\pi}\big).\end{align}
where $|u_k^{\pm}\rangle =\frac{1}{\sqrt{2}}\big(\pm e^{-i\theta_k}|A\rangle+|B\rangle\big)$ is the cell periodic Bloch state. We have $\gamma_{\pm}=\pi~(0)$ for $v/w<1~(>1)$ indicating a topologically non-trivial (trivial) phase of the waveguide. We show below how $\theta_k$ qualitatively differentiates quantum interference in topologically trivial and non-trivial phase for a single photon propagating through a giant 2LE. 

\begin{figure*}
\centering
\begin{subfigure}[b]{0.47\textwidth}
\includegraphics[width=\textwidth]{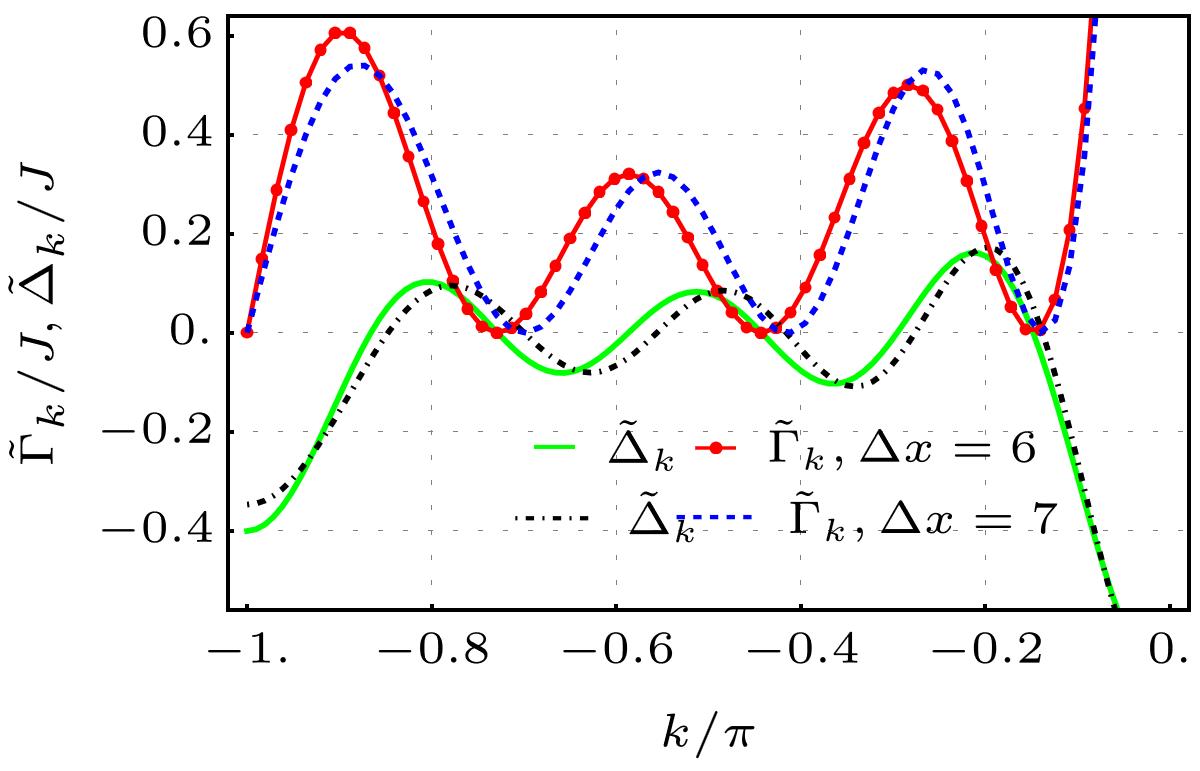}
\caption{SSH: $A$-$B$}
\end{subfigure} 
\begin{subfigure}[b]{0.47\textwidth}
\includegraphics[width=\textwidth]{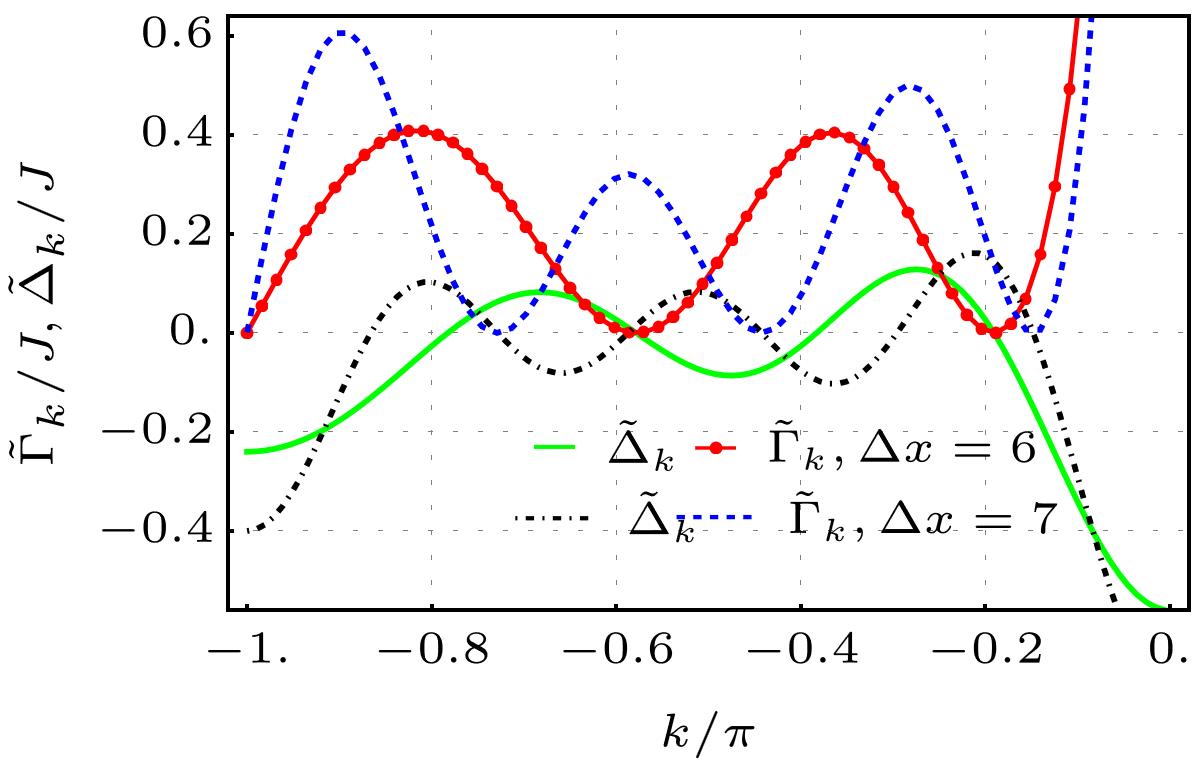}
\caption{SSH: $B$-$A$}
\end{subfigure}
\caption{Wave vector $(k)$ dependence of decay rate $\tilde{\Gamma}_k$ and Lamb shift $\tilde{\Delta}_k$ for (a) $A$-$B$ and (b) $B$-$A$ configuration of a giant 2LE side coupled to the SSH waveguide  at two distinct sites with an odd (even)  unit cell separation  $\Delta x$ in the trivial, $\delta=-0.5$  (non-trivial, $\delta=0.5$) phase of the waveguide. Other parameters, $g_1/J=g_2/J=0.2$. }\label{SSHgiant}
\end{figure*}

We again consider a 2LE is side coupled to two sites of the SSH waveguide. These two sites can be both $A$ or both $B$ or one $A$ and another $B$ sublattice. Our studied constraint on the decay width of a giant 2LE is only possible for a giant emitter coupled to two sublattice sites of $A$ and $B$ as explained below. Interestingly, it has been shown recently that particle-hole symmetry is satisfied by $A$-$A$ or $B$-$B$ coupling but not by $A$-$B$ or $B$-$A$ coupling \cite{PhysRevA.106.033522}. Below, we discuss separately two configurations of one $A$ and another $B$ coupling as depicted for smallest separation in Fig.~\ref{G2LESSH}. 

{\it{$A$-$B$ type configuration}}:  A 2LE is connected to sublattice $A$ of $0^{\text{th}}$ unit cell and sublattice $B$ of $\Delta x^{\text{th}}$ unit cell of an SSH waveguide. The coupling amplitudes at these two connection points $(0\,A,\Delta x\,B)$  are $g_1$ and $g_2$, respectively. We take $\Delta x\geq 0$. Thus, an incoming photon from the left of the emitter encounters the emitter first when it reaches sublattice $A$ of $0^{\text{th}} $ unit cell. The light-emitter interaction Hamiltonian within the RWA is written as,
\begin{align}
\hat{H}_{iAB}=g_1\big(a_{0}^{\dagger}\sigma+\sigma^{\dagger}a_{0}\big)+g_2\big(b_{\Delta x}^{\dagger}\sigma+\sigma^{\dagger}b_{\Delta x}\big). \label{IAB}
\end{align}
We consider an right-moving single-photon input state $|\phi_k^{+}\rangle$, with an energy in the upper band, $E=\omega_k^{+}$, for $k\in [-\pi,0]$, and the emitter in its ground state. The range of $k\in [-\pi,0]$ ensures a positive group velocity in the upper band, i.e., $v_g^+(k)=\frac{\partial\omega_k^{+}}{\partial k}=-vw\sin{k}/|v+we^{ik}|>0$. Since $\hat{H}_{iAB}$ conserves total excitation of light and emitter, an eigenstate of the full Hamiltonian $\hat{H}_{SSH}+\hat{H}_e+\hat{H}_{iAB}$ in the single-excitation sector can be written as
\begin{align}
    |\psi_k\rangle=\sum_x (\psi_k(x,A)a_x^{\dagger}+\psi_k(x,B)b_x^{\dagger})|\varphi,g\rangle+\psi_e |\varphi,e\rangle,\label{sEx}
\end{align} 
where $\psi_k(x,\alpha)\equiv \big(\langle g|\langle x,\alpha|\big)|\psi_k\rangle$ ($\alpha=A,B$) and $\psi_e$ are the probability amplitudes of a single photon in the waveguide and an excited emitter, respectively. Plugging Eqs.~\ref{sEx},\ref{IAB} in the LS equation (\ref{LippmannRetardG}), we get the following relations:
\begin{align}
\psi_k(x,\alpha)=\phi_k^+(x,\alpha)+&\Big(g_1 \langle g|\langle x,\alpha|G_o^{R}|0,A \rangle|g \rangle\nonumber \\&+g_2  \langle g|\langle x,\alpha|G_o^{R}|\Delta x,B \rangle|g \rangle \Big)\psi_e, \label{AB SSH Lippmann}\\
\psi_e  = \frac{1}{E-\omega_e}\Big(g_1 \psi_k(0&, A)+g_2 \psi_k(\Delta x, B)\Big).
\end{align}
These two equations can be solved for an initial state $\phi_k^+(x,\alpha) \equiv  \langle x,\alpha | \phi_k^+\rangle $, using the Green's functions evaluated in Appendix~\ref{AppendixA}. Thus, we find the following wavefunction away from the scattering region:
\begin{align} 
&  \psi_k(x,A)=\begin{cases}
  \frac{1}{\sqrt{4\pi}}\big(e^{ikx}e^{-i\theta_k}+ r_k e^{-ikx}e^{i\theta_k}\big)\quad\text{for $x<0$,} \\
   \frac{1}{\sqrt{4\pi}} t_k e^{i k x}e^{-i\theta_k}\quad\quad\quad\quad\quad\quad\text{for $x>\Delta x$,} \\
   \end{cases} \label{scattering eigenstate in A SSH} \\
 & \psi_k(x,B)=\begin{cases}
     \frac{1}{\sqrt{4\pi}}\big(e^{ikx}+ r_k e^{-i k x}\big)\quad\quad &\text{for $x<0$,} \\
   \frac{1}{\sqrt{4\pi}} t_k e^{ikx}\quad\quad &\text{for $x>\Delta x$,}
   \end{cases} \label{scattering eigenstate in B SSH}
\end{align}
where $r_k$ and $t_k$ are the reflection and transmission amplitudes. The transmission and reflection coefficient, $\tilde{T}_k=|t_k|^2, \tilde{R}_k=|r_k|^2$, are 
\begin{align}
      \tilde{T}_k=1-\tilde{R}_k,\;\tilde{R}_k=\frac{(\tilde{\Gamma}_k/2)^2}{(\omega_k^{+}-\omega_e-\tilde{\Delta}_k)^2+(\tilde{\Gamma}_k/2)^2},\label{AB reflection coeff SSH}
\end{align}
which have similar form of those for the TB waveguide (\ref{reflection coeff TB}). The $k$-dependent Lamb shift $\tilde{\Delta}_k$ and total decay width $\tilde{\Gamma}_k$ for photons in the upper band are:
\begin{align}
    \tilde{\Delta}_k[g_1,g_2]&=\frac{g_1 g_2}{v_g^+(k)}\sin{\big(k\Delta x+\theta_k\big)},\label{ABDelta} \\
    \tilde{\Gamma}_k[g_1,g_2]&=\frac{1}{v_g^+(k)}\Big(g_1^2+g_2^2+2g_1g_2\cos{\big(k\Delta x+\theta_k\big)}\Big).\label{ABGamma}
\end{align}
These expressions (\ref{ABDelta},\ref{ABGamma}) for the SSH waveguide are mostly similar to those (\ref{TBeshift},\ref{TBdecay}) for the TB waveguide apart from the appearance of an extra $\theta_k$ factor in the sinusoidal dependence. The form of photon group velocity is different for the two structural waveguides, and $v_g^+(k)$ for the SSH waveguide depends also on $\theta_k$. Similar to our previous discussion for the TB waveguide, we get here a condition for zero interference in $\tilde{\Gamma}_k[g_1,g_2]$ as $(v/w)\cos{k\Delta x}=-\cos{k\big(\Delta x+1\big)}$, which admits a real solution of $k \in (-\pi,0)$ for $\Delta x \neq 0$ both in topological and trivial phases of the SSH waveguide. Therefore, the total decay width of the giant 2LE coupled to a SSH waveguide can become a sum of the individual decay width for some incident wavevector when $\Delta x \neq 0$ for all $v,w$. Nevertheless, the above condition needs special attention for $\Delta x =0$, when it becomes $\cos{k}=-v/w$, which has no physical solution for real $k$ values when $v>w$ in the trivial phase of the SSH waveguide. Since $\cos{k}=-v/w$ has a physical solution for real $k$ values when $v<w$, we can get $\tilde{\Gamma}_k[g_1,g_2]=\tilde{\Gamma}_k[g_1,0]+\tilde{\Gamma}_k[0,g_2]$ in the topological phase.

The photon group velocity $v_g^+(k)$ for SSH waveguides also tends towards zero near the band edges at $k=0,\pm\pi$. From Eqs.~\ref{ABGamma}, \ref{ABDelta}, we observe that $\tilde{\Gamma}_k[g_1,g_2]=0$ implies $\cos{(k\Delta x+\theta_k)}=-1$, i.e., $k_n=-((2n+1)\pi+\theta_k)/\Delta x$ where $n=0,1,2,\dots$ for $g_1=g_2$. Now, $\theta_k=0$ at $k=0,\pm \pi$ in the trivial phase, and $\theta_k=0,\pm\pi$ at $k=0,\pm \pi$, respectively, in the non-trivial phase. Thus, for the trivial phase of a SSH waveguide, $\tilde{\Delta}_k[g_1,g_1]$ vanishes at the above $k_n$ values excluding $k_n=-\pi$ for odd integer values of $\Delta x$, which is similar to the TB waveguide. However, contrary to the TB waveguide, we get $\tilde{\Gamma}_{k \to -\pi}[g_1,g_1]=0$ and a non-zero $\tilde{\Delta}_{k \to -\pi}[g_1,g_1]$ for even integer values of $\Delta x$ in the non-trivial phase of a SSH waveguide due to $\theta_{k=-\pi}=-\pi$. Therefore, both for an odd and an even integer unit cells separation between the two couplings  of a giant 2LE with an SSH waveguide, we have $\tilde{\Gamma}_{k \to -\pi}[g_1,g_1]=0$ and $\tilde{\Delta}_{k \to -\pi}[g_1,g_1]=(-1)^{\Delta x} g_1^2((v-w)\Delta x-w)/(vw)$, respectively for the trivial and non-trivial phase. We display these features in Fig.~\ref{SSHgiant}(a).

 {\it{$B$-$A$ type configuration}} : We next consider a reverse configuration in which a 2LE is connected to $(0\,B,\Delta x\,A)$  with coupling strengths $g_1,g_2$, respectively and $\Delta x> 0$. An incoming photon from the left of the giant emitter first meets the 2LE at sublattice $B$ of $0^{\text{th}}$ unit cell. The light-emitter interaction Hamiltonian within the RWA is written as,
\begin{align}
    \hat{H}_{iBA}=g_1\big(b_{0}^{\dagger}\sigma+\sigma^{\dagger}b_{0}\big)+g_2\big(a_{\Delta x}^{\dagger}\sigma+\sigma^{\dagger}a_{\Delta x}\big). \label{IBA}
\end{align}
 The form of the Lamb shift and collective decay width for this configuration is mostly the same as those (\ref{ABDelta},\ref{ABGamma}) in $A$-$B$ type configuration apart from a sign change for $\theta_k$ factor, i.e., $\theta_k$ is replaced by $-\theta_k$.  The condition for fully suppressing the interference in $\tilde{\Gamma}_k$ for this configuration is then $(v/w)\cos{k\Delta x}=-\cos{k(\Delta x-1)}$, which admits a real physical solution of $k \in (-\pi,0)$ for both topological and trivial phase of the SSH waveguide when $\Delta x \neq 1$. Nevertheless, when $\Delta x=1$, the above equation reads $\cos{k}=-w/v$, which does not have a real solution for $k$ in the topological phase for $v<w$. Thus, we proof a constraint (no-go theorem) on the collective decay width $\tilde{\Gamma}_k$ of the giant 2LE in one of the two topological phases for these two configurations. 

The above discussed features of a decoherence-free strong coupling indicating a large Lamb shift and zero collective decay width around the band edges for $A$-$B$ configuration hold true for $B$-$A$ configuration. However, the value of Lamb shift around the band edges has changed, and it is given by $\tilde{\Delta}_{k \to -\pi}[g_1,g_1]=(-1)^{\Delta x} g_1^2((v-w)\Delta x+w)/(vw)$, which is higher (lower) for $B$-$A$ configuration than $A$-$B$ configuration for an odd (even) number of unit cells separation in the trivial (non-trivial) phase. We show this comparison in Fig.~\ref{SSHgiant}(a) and \ref{SSHgiant}(b).
\begin{figure}
\includegraphics[height=3.0cm,width=8.5cm]{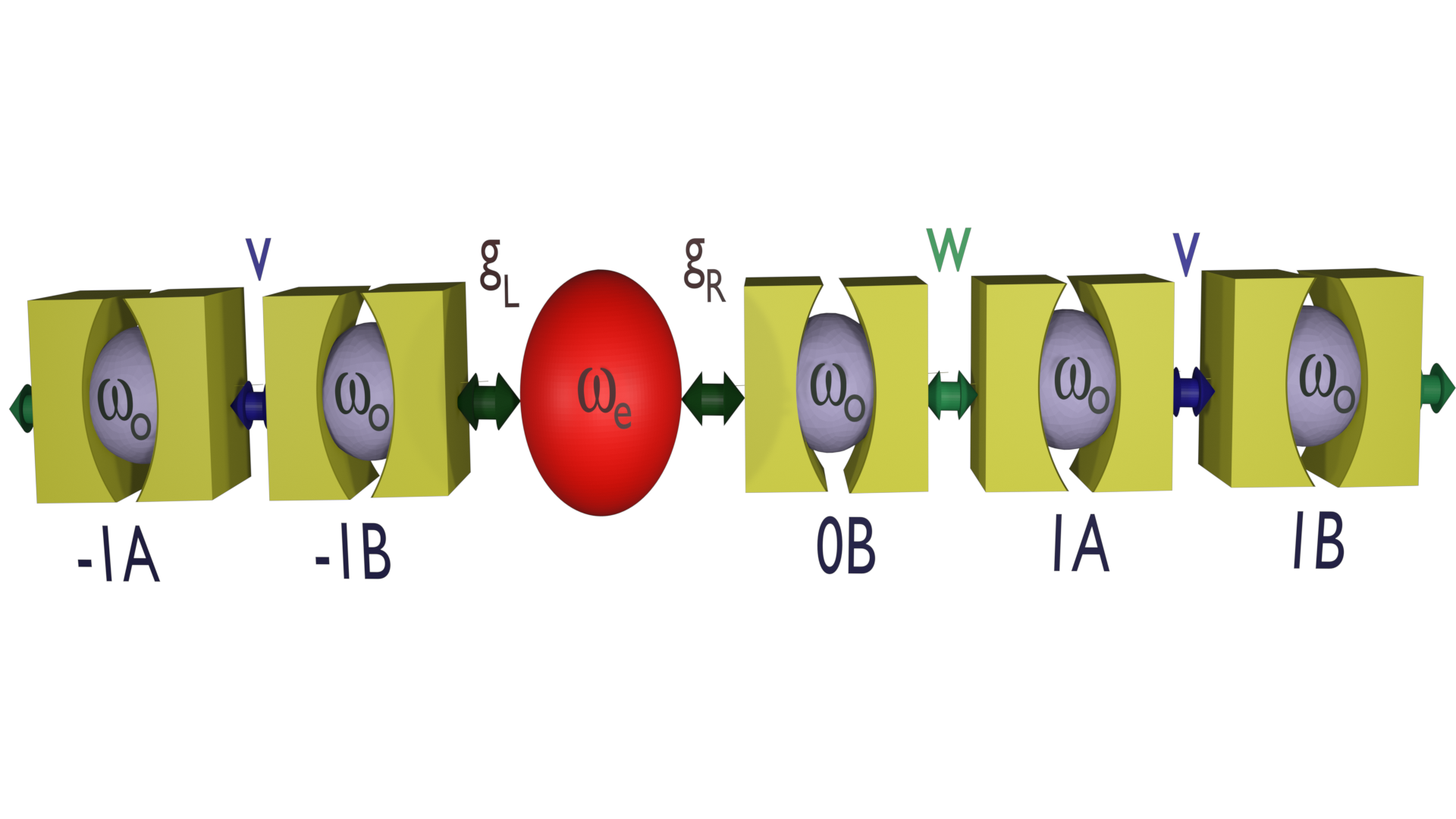}
\caption{A 2LE with a transition frequency $\omega_e$ is direct coupled to an SSH waveguide by replacing the sublattice site at $(x,\alpha)= (0,A)$. The intra- (inter-) unit cell hopping are $v~(w)$ and the frequency of each sublattice site is $\omega_0$. The coupling amplitudes of the 2LE to the sublattice site $B$ of unit cells $x=-1,1$ are $g_L,g_R$.}
\end{figure}

\section{Direct-coupled vs. side-coupled emitter}\label{analogy}
For unstructured (continuous) waveguides with a linear energy-momentum dispersion and an infinite bandwidth (i.e., no band edge) for propagating photons, the transport properties of a side-coupled emitter are closely related to those of a direct-coupled emitter. Then, the single-photon transmission and reflection amplitude for a side-coupled 2LE are identical to the single-photon reflection and transmission amplitude, respectively, for a direct-coupled 2LE \cite{PhysRevA.79.023837}. We here examine such a comparison for structured waveguides with a finite band width. An emitter directly coupled to waveguides requires coupling the emitter at the edges of the waveguides \cite{mirhosseini2018superconducting,faraon2007efficient}. 

\subsection{TB waveguide : Lamb shift}
The Hamiltonian of a 2LE direct coupled to the TB waveguides is 
\begin{align}\label{directTBHam}
\hat{H}_{dTB} =& \sum_{x\ne-1,0}J\big(c_{x}^{\dagger} c_{x+1}+ c_{x+1}^{\dagger}c_{x} \big)+\sum_{x\ne 0}\omega_o c_{x}^{\dagger} c_{x}+\hat{H}_e \nonumber \\ &+ (g_L c_{-1}^{\dagger}+g_R c_{1}^{\dagger})\sigma + \sigma^{\dagger}(g_L c_{-1}+g_R c_{1}),
\end{align}
where, the sums over site $x$ are redefined to insert a 2LE at site $x=0$. The 2LE is direct coupled to sites $x=-1, 1$ of the TB waveguide with strength $g_L$ and $g_R$, respectively. The Hamiltonian $\hat{H}_{TB}+\hat{H}_e+\hat{H}_i$ in Eqs.~\ref{eq:TB hamiltonian},\ref{TLS Hamiltonian},\ref{intGTB} for $g_2=0$ is the side-coupled analog of Eq.~\ref{directTBHam} when $g_L=g_R=g_1/2$ \cite{PhysRevA.83.043823}. The single-photon transmission and reflection coefficient of the analog side-coupled 2LE can be obtained from Eq.~\ref{reflection coeff TB} by setting $\Delta_k=0$ and $\Gamma_k=-g_1^2/(J\sin k)$ for $g_2=0$.

We derive the single-photon transmission and reflection coefficient $T_d,R_d$ for an incident photon from the left of the direct-coupled 2LE:
\begin{align}
    T_d=\frac{\Gamma_L\Gamma_R}{(\omega_k-\omega_e-\Delta_d)^2+\frac{(\Gamma_L+\Gamma_R)^2}{4}}, R_d=1-T_d,\label{transDTB}
\end{align}
where the decay widths $\Gamma_L=-(2 g_L^2\sin k)/J,\; \Gamma_R=-(2 g_R^2\sin k)/J$ and the Lamb shift $\Delta_d=(g_L^2+g_R^2)\cos k/J$. We thus find a finite Lamb shift for a 2LE direct coupled to structured waveguides, and $\Delta_d$ survives even when either $g_L$ or $g_R$ is zero \cite{mirhosseini2018superconducting}. However, no Lamb shift appears for a 2LE side coupled to a TB waveguide at one site. Therefore, unlike unstructured waveguides with an infinite bandwidth, the single-photon transmission and reflection coefficient for a side-coupled 2LE are not the same to the single-photon reflection and transmission amplitude for a direct-coupled 2LE for the TB waveguide. 

We further notice that the total decay width of a side-coupled 2LE and that of a direct-coupled 2LE are very different in their form for a TB waveguide. The total decay width are $\Gamma_d=\Gamma_R+\Gamma_L=-(g_1^2 \sin k)/J$ and $\Gamma_s=-g_1^2/(J \sin k)$, respectively, for a direct-coupled 2LE and a side-coupled 2LE when $g_L=g_R=g_1/2$. While $\Gamma_d$ is largest near $k=-\pi/2$ and vanishes near the band edges at $k=0,-\pi$, $\Gamma_s$ shows a very opposite trend at these quasi-momenta. Further, the Lamb shift is maximum and the decay width is zero near the band edges for a direct-coupled 2LE as observed by \textcite{mirhosseini2018superconducting}. As expected, the differences of the total decay width and the Lamb shift between a 2LE direct coupled and side coupled to the TB waveguide disappear in the continuum limit at the matching condition for the wave vector $lk \sim -\pi/2$ where we set the lattice constant $l=1$.

\subsection{SSH waveguide}\label{dSSH}
Next, we explore a 2LE direct coupled to SSH waveguides and discuss a new feature of asymmetric dependence of the Lamb shift on the light-matter couplings in such a system that emerges due to topological features of the SSH waveguides. The Hamiltonian of the full system is  
\begin{align}
   \hat{H}_{dSSH} =&\sum_{x \ne 0} \omega_o a_{x}^{\dagger} a_{x}+ \sum_{x} \omega_o b_{x}^{\dagger} b_{x}+ \sum_{x \ne 0} v\big(a_{x}^{\dagger} b_{x}+ b_{x}^{\dagger} a_{x}\big) \nonumber \\ &+ \sum_{x \ne -1} w \big(a_{x+1}^{\dagger} b_{x}+ b_{x}^{\dagger} a_{x+1}\big)+\hat{H}_e \nonumber \\ &+(g_L b_{-1}^{\dagger}+g_R b_{0}^{\dagger})\sigma + \sigma^{\dagger}(g_L b_{-1}+g_R b_{0}), \label{dSSHHamiltonian}
\end{align}
where, the sums over unit cell $x$ from $-\infty$ to $\infty$ are redefined to insert a 2LE at sublattice site $A$ of $x=0$ unit cell. The 2LE is direct coupled to sublattice site $B$ of $x=-1, 0$ unit cells of the SSH waveguide with strength $g_L$ and $g_R$, respectively. The Hamiltonian $\hat{H}_{SSH}+\hat{H}_e+\hat{H}_{iAB}$ in Eqs.~\ref{SSH Hamiltonian},\ref{TLS Hamiltonian},\ref{IAB} for $g_2=0$ is the side-coupled analog of Eq.~\ref{dSSHHamiltonian} when $g_L=g_R=g_1/2$. The single-photon transmission and reflection coefficient of the analog side-coupled 2LE can be found from Eq.~\ref{AB reflection coeff SSH} by setting $\tilde{\Delta}_k=0$ and $\tilde{\Gamma}_k=g_1^2/(v_g^+(k))\equiv \tilde{\Gamma}_s$ for $g_2=0$.
 \begin{figure}
\centering
\includegraphics[width=0.47\textwidth]{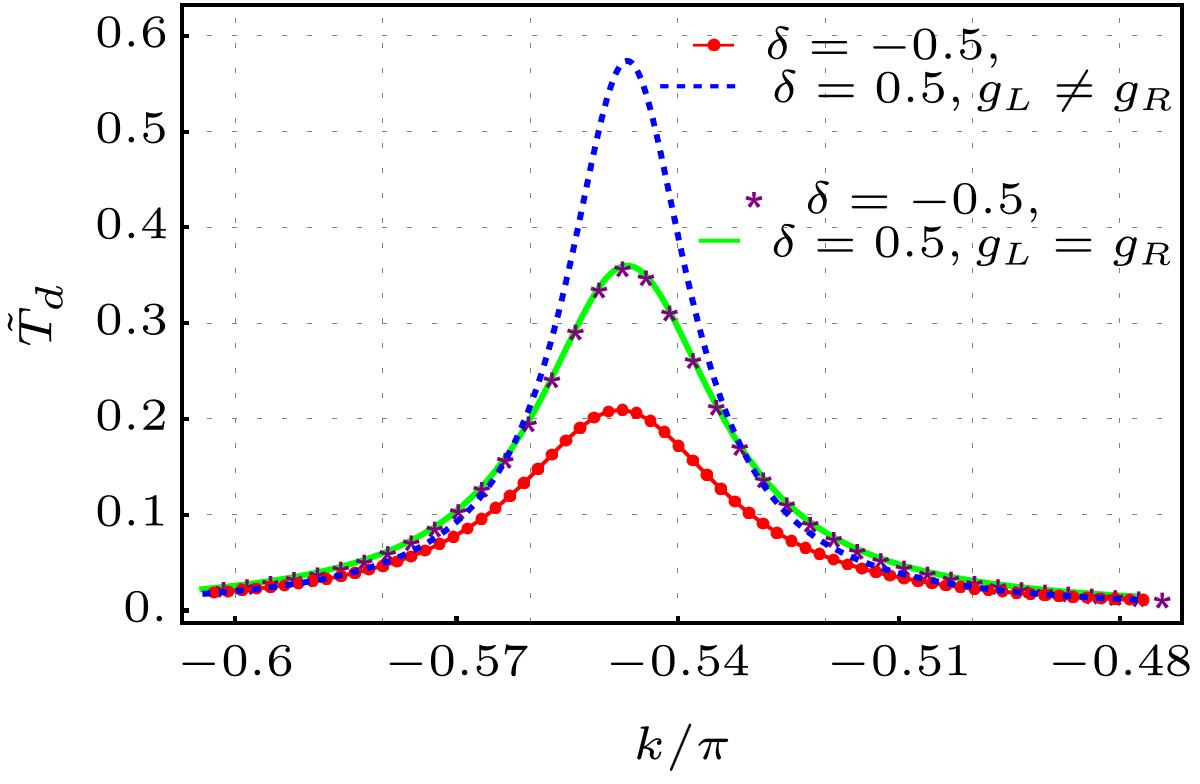}
\caption{Single-photon transmission coefficient $\tilde{T}_d$ through a direct-coupled 2LE for topologically trivial and non-trivial phase of the SSH waveguide. While $\tilde{T}_d$ remains the same in the both phases when $g_L=g_R~(=0.1J)$, its value changes significantly between the two phases when $g_L\ne g_R$, e.g., $g_L/J=0.08, g_R/J=0.11$. The other parameters $\omega_o/J=3.0$ and $\omega_e/J=4.5$.}\label{SSHdtrans}
\end{figure}

The form of the single-photon transmission and reflection coefficient, $\tilde{T}_d, \tilde{R}_d$, for an incident photon from the left of the direct-coupled 2LE in Eq.~\ref{dSSHHamiltonian} is the same as those for the direct-coupled model of the TB waveguide in Eq.~\ref{transDTB} after replacing $\omega_k$ by $\omega_k^{+}$ for the upper band. However, the decay widths $\tilde{\Gamma}_L, \tilde{\Gamma}_R$ and the Lamb shift $\tilde{\Delta}_d$ have  exciting dependence on the inter-cell and intra-cell hopping amplitudes and the emitter-waveguide coupling strengths.
\begin{align}
    &\tilde{\Gamma}_L=\frac{-2g_L^2 v \sin k}{w f_k},~ \tilde{\Gamma}_R= \frac{-2g_R^2 w \sin k}{v f_k},\label{transDSSH0} \\
    &\tilde{\Delta}_d= \frac{g_L^2(w+v \cos k)}{w f_k}+ \frac{g_R^2(v+w \cos k)}{v f_k},\label{transDSSH}
\end{align}
which show that the decay widths depend asymmetrically on $v,w$, i.e., $\tilde{\Gamma}_L$ does not transform to $\tilde{\Gamma}_R$ when $g_L$ is replaced by $g_R$. Further, $\tilde{\Delta}_d$ also depends asymmetrically on $g_L$ and $g_R$ due to the appearance of $v,w$. Nevertheless, $\tilde{\Gamma}_L$ transforms to $\tilde{\Gamma}_R$ when the couplings $g_L,g_R$ and the hoppings $v,w$ are simultaneously exchanged. $\tilde{\Delta}_d$ remains invariant under exchange of $g_L$ with $g_R$ and $v$ with $w$. Therefore, the single-photon transport coefficients $\tilde{T}_d$ and $\tilde{R}_d$ remain the same for an incident photon from the left or the right side of the 2LE even when $g_L \ne g_R$.  Thus, there is no rectification of a single photon in these direct-coupled spatially asymmetric models as expected following \cite{PhysRevB.81.155117}. However, the value of $\tilde{T}_d$ changes significantly for $g_L \ne g_R$ when $v,w$ are exchanged which we show in Fig.~\ref{SSHdtrans}. Further, a single photon can fully transmit through this direct-coupled 2LE system only when $\tilde{\Gamma}_L=\tilde{\Gamma}_R$.  
\begin{figure}
\centering
\includegraphics[width=0.47\textwidth]{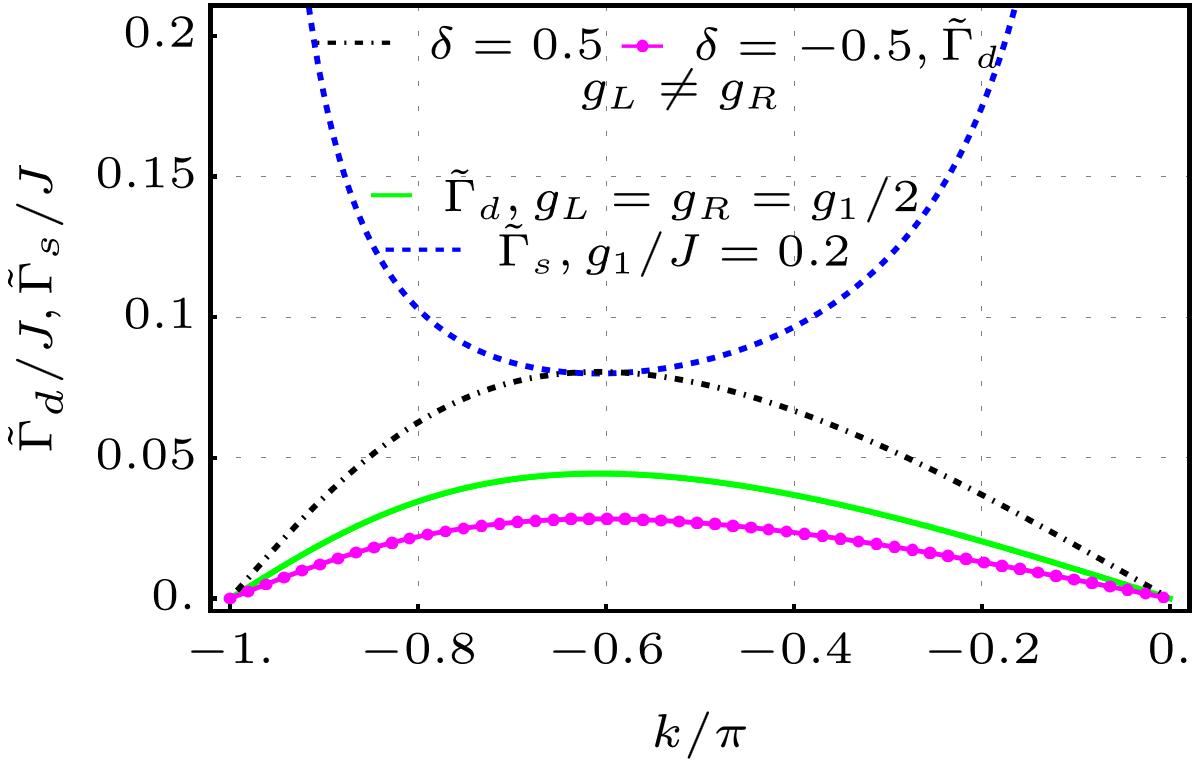}
\caption{Comparison between decay rates $\tilde{\Gamma}_d,\tilde{\Gamma}_s$ of a direct-coupled (green full line) and a side-coupled (blue dashed line) 2LE to the SSH waveguide when $g_L=g_R=g_1/2=0.1J$, $|\delta|=0.5$. The black dot-dashed and magenta dot lines show decay rates $\tilde{\Gamma}_d$ of a direct-coupled 2LE in different topological phases of the SSH waveguide when coupling amplitudes are unequal i.e., $g_L/J=0.07,g_R/J=0.14$. }\label{SSHddecay}
\end{figure}
\begin{figure*}
\includegraphics[height=3.8cm,width=16.5cm]{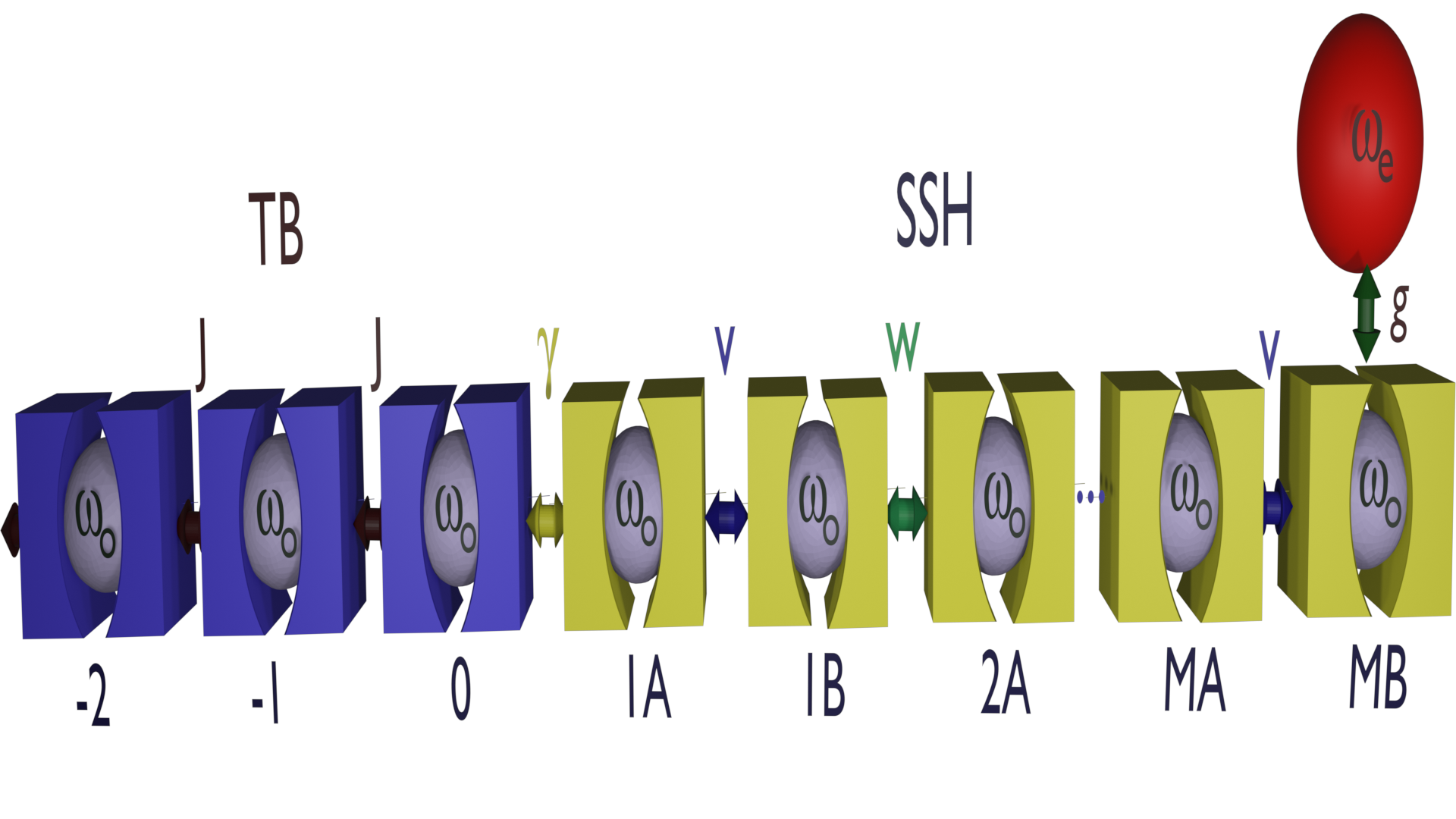}
\caption{A heterojunction of a long TB waveguide and a finite SSH waveguide is connected to a 2LE at one end with an amplitude $g$. Here, $J$ is the TB hopping, $v(w)$ is the intra-(inter-) cell hopping of the SSH part, and each site of the heterojunction modeled as a resonator of frequency $\omega_o$. $\gamma$ is the tunneling rate at the junction between the TB and SSH parts. }
\label{Switch}
\end{figure*}
The analogy between transport coefficients of side-coupled and direct-coupled models for $g_L=g_R=g_1/2$ also breaks down for SSH waveguides due to the finite Lamb shift appearing in a direct-coupled 2LE. Thus, the single-photon transmission and reflection coefficient for a side-coupled 2LE do not match with the single-photon reflection and transmission amplitude for a direct-coupled emitter both for TB and SSH waveguides. Like the TB waveguide case, $\tilde{\Delta}_d$ is again non-zero even for one-sided coupling (i.e., for $g_L=0$ or $g_R=0$), indicating the emergence of a finite Lamb shift when an emitter is connected to the end of a structured waveguide with a finite bandwidth \cite{mirhosseini2018superconducting}. We prove the previous conclusion explicitly in Appendix~\ref{AppendixB} by deriving the Lamb shift and the decay width of a 2LE connected to a semi-infinite SSH waveguide at one end. The Appendix~\ref{AppendixB} also clarifies the contributions of the edge modes of the SSH chains to the Lamb shift. The decay width $\tilde{\Gamma}_d=\tilde{\Gamma}_L+\tilde{\Gamma}_R$ of the direct-coupled 2LE to SSH waveguides has a very different form and quasi-momentum dependence from those of $\tilde{\Gamma}_s$ of a side-coupled 2LE to the SSH waveguide, which we demonstrate in Fig.~\ref{SSHddecay}.  In Fig.~\ref{SSHddecay}, we further depict the topology dependence of $\tilde{\Gamma}_d$ when $g_L \ne g_R$.

\section{Heterojunction of structured waveguides: quantum switch}\label{Qswitch}

We finally study the dynamics of a single photon in a heterojunction made of a long TB waveguide and a short SSH waveguide. Let us consider a 2LE being connected to the one end of the heterojunction as depicted in Fig.~\ref{Switch}. We show below that a photon from the excited emitter can propagate to the TB waveguide within a finite time only when the SSH waveguide is in the topological phase. Thus, the whole system acts as a quantum switch. We choose the TB waveguide of $N+1$ sites and the SSH waveguide of $M$ unit cells in the region of $x \in [-N,0]$ and $x\in [1,M]$ of the heterojunction, respectively. 
The total Hamiltonian of the heterojunction is $\hat{H}_{TB}+\hat{H}_{SSH}+\hat{H}_{\gamma}$, where $\hat{H}_{\gamma}$ denotes tunnel coupling between the TB and SSH waveguides with a rate $\gamma$: 
\begin{align}
    \hat{H}_{\gamma}=\gamma \big(c_{0}^{\dagger} a_{1}+a_{1}^{\dagger} c_{0}\big).
\end{align}
The 2LE is coupled to the last sublattice site, i.e., $(x,\alpha)=(M,B)$ of the SSH waveguide. We fix the size $M$ of the SSH waveguide such that there is a finite coupling between the mid-gap edge modes at the boundaries of the SSH waveguide in the topological phase. Thus, $M \le 1/\log[w/v]$. We further consider the transition frequency $\omega_e$ of the 2LE around the middle of the SSH spectrum, i.e., $\omega_e=\omega_o$. The last ensures a finite overlap between the edge modes of the SSH waveguide and the 2LE. We again write the coupling of the 2LE with the heterojunction within the RWA:
\begin{align}
    \hat{H}_{I}=g\,\big(b_{M}^{\dagger}\sigma^{-}+ \sigma^{+}b_{M}\big),
\end{align}
where $g$ is the coupling amplitude. 

We calculate the time evolution of a single-excitation initial state by numerically solving the time-dependent Schr{\"o}dinger equation, $i\partial_t|\psi(t)\rangle=\hat{H}_T |\psi(t)\rangle$,  with the total Hamiltonian, $\hat{H}_T=\hat{H}_{TB}+\hat{H}_{SSH}+\hat{H}_{\gamma}+\hat{H}_{e}+\hat{H}_{I}$. A general wave function $|\psi(t)\rangle$, at any time $t$, for the whole system in the single-excitation sector can be written as,
\begin{align}
    |\psi(t)\rangle=&\big(\sum_{x=-N}^{0}\psi_x(t) c_x^{\dagger}+\sum_{x=1}^{M}( \psi_{xA}(t) a_x^{\dagger} +\psi_{xB}(t)b_x^{\dagger})\nonumber\\ &+\psi_e(t)\sigma^{+}\big)|\varphi,g\rangle.
\end{align}
 We cast the time-dependent Schr{\"o}dinger equation into a first-order discrete time difference equation with small time steps (e.g., $\Delta t=10^{-3}$) and determine the probability amplitudes $\psi_x,\psi_{x A},\psi_{x B},\psi_e$ up to a future time $t=t_{\rm f}$ for the initial condition of an excited emitter with no photon in the heterojunction, i.e., $|\psi(t=0)\rangle=|\varphi, e\rangle$. The simulation run-time $t_{\rm f}$ is adequately fixed to avoid the boundary scattering at the end of the TB waveguide for a finite $N$, i.e., $v_g(k)t_{\rm f}<N$ where $v_g(k)$ is the group velocity in a TB waveguide and wave-vector $k$ corresponds to the energy of the excitation traveling in TB waveguide.  We calibrate $t_{\rm f}$ to get a good quantum switch behavior for some parameter sets of our interest. 

 \begin{figure}
\includegraphics[width=0.45\textwidth]{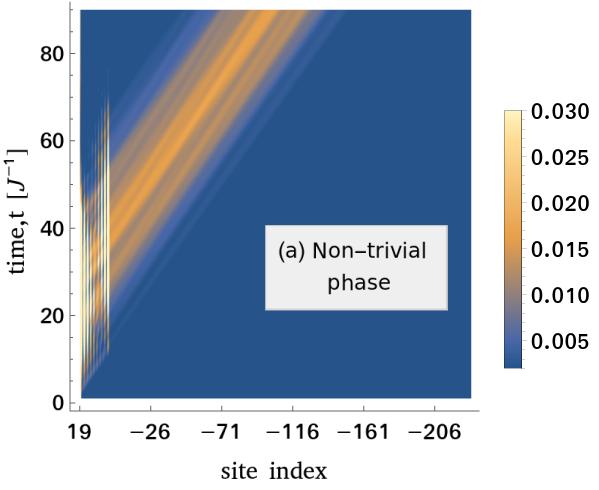}
\includegraphics[width=0.45\textwidth]{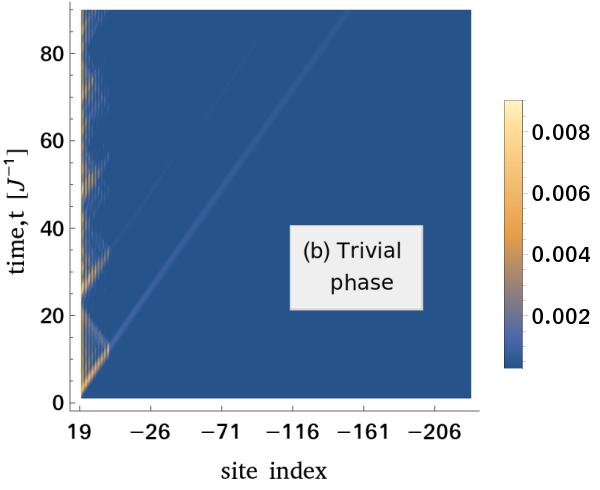}
\caption{Spatio-temporal evolution of single excitation in the heterojunction from an initially excited 2LE for (a) non-trivial ($\delta=-0.5$), and (b) trivial ($\delta=0.5$)  phase of the SSH waveguide. Color legends indicate values of single-excitation probability. While the 2LE and the SSH waveguide are labelled by site index $19$ and from $18$ to $1$, respectively, the rest (between $0$ to $-229$) represents the TB waveguide. Other parameters are $g/J=0.07$, $\gamma/J=0.4$, $\omega_o/J=3.0$, $t_{\rm f}=110/J$, $\delta t=0.001/J$, $N=229$,  $M=9$. }\label{timeEv} 
\end{figure}

\begin{figure*}
\centering
\begin{subfigure}[b]{0.47\textwidth}
\includegraphics[width=\textwidth]{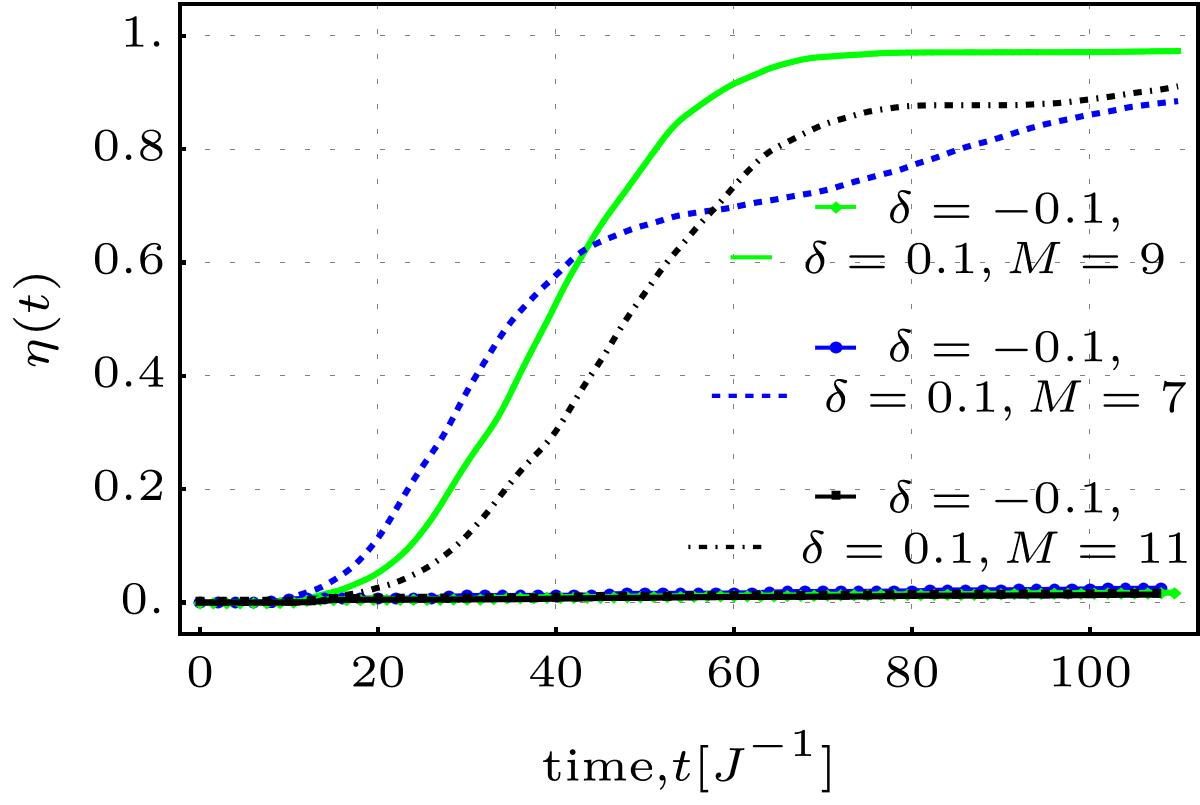}
\end{subfigure}
\begin{subfigure}[b]{0.47\textwidth}
\includegraphics[width=\textwidth]{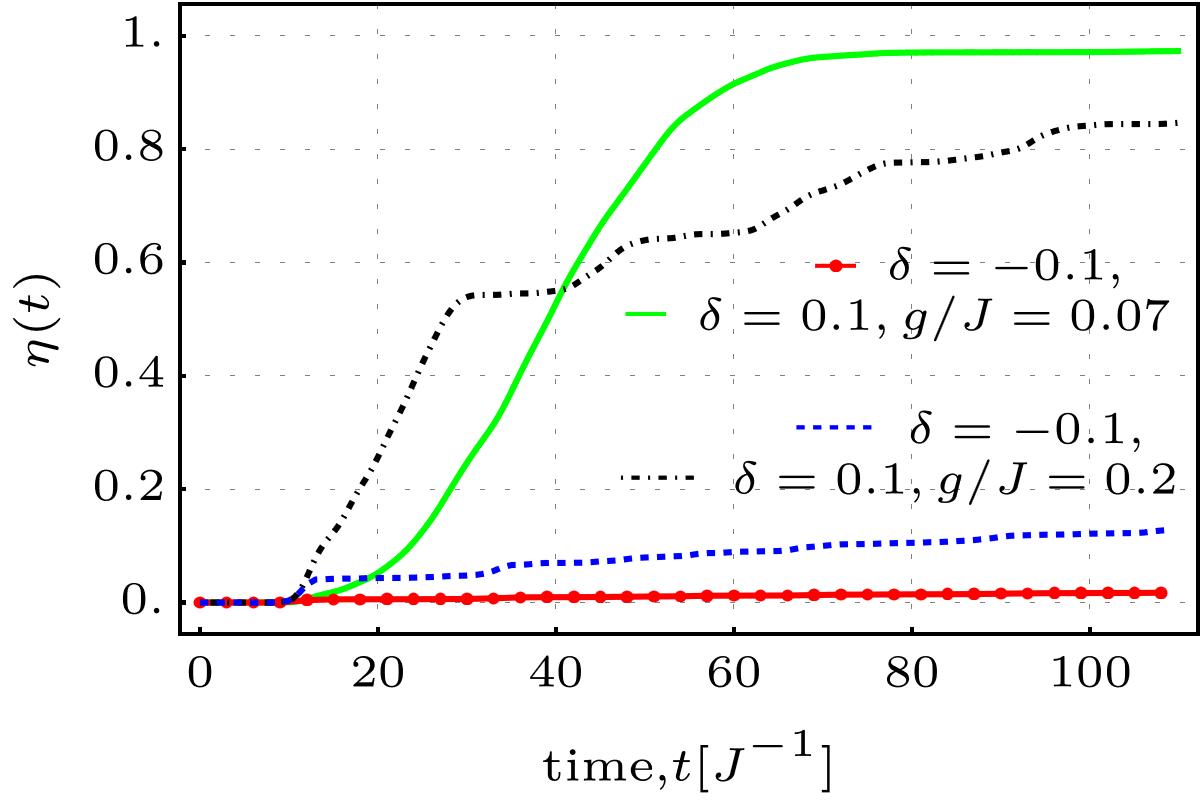}
\end{subfigure}
\caption{Dependence of photon emission efficiency $\eta(t)$ on the SSH length $M$ (a) and the emitter's coupling amplitude $g$ (b) in the topologically trivial ($\delta<0$) and non-trivial ($\delta>0$) phase of the SSH waveguide. Other parameters are $\gamma/J=0.4$, $\omega_o/J=3.0$, $N=219$, $t_{\rm f}=110/J$, $\delta t=0.001/J$ in both plots, and $g/J=0.07$ for (a) and $M=9$ for (b). }\label{efficiency}
\end{figure*}

In Fig.~\ref{timeEv}, we show our simulation results for the time evolution of a single excitation over the sites of the heterojunction and the emitter. For the trivial phase of the SSH waveguide in Fig.~\ref{timeEv}(b), we observe that the excitation is mainly confined within the emitter and the SSH waveguide with a slight leakage to the TB waveguide within our observation time $t_{\rm f}$. On the other side, we find a significant leakage of a single photon to the TB waveguide within the time $t_{\rm f}$ for the SSH waveguide in the topologically non-trivial phase, as shown in Fig.~\ref{timeEv}(a). Therefore, the hybrid system acts as a quantum switch allowing a single photon to transmit to the TB waveguide from an excited emitter depending on the topological nature of the SSH waveguide. The heterojunction has two edge modes at the boundaries of the SSH waveguide near the energy $\omega_0$ in the topological phase. These modes are strongly hybridized with the 2LE, as we explain below. 

The edge mode around $x=1$ unit cell is further coupled by $\hat{H}_{\gamma}$ to continuum band modes of the TB waveguide. Thus, there is a strong propagation channel from the emitter to the TB waveguide via the edge modes for the SSH waveguide in the topologically non-trivial phase. Due to the absence of edge modes in the trivial phase, such a propagation channel is mostly missing, and the emission from the 2LE hardly leaks to the TB waveguide within $t_{\rm f}$.

Finally, we point out the role of different parameters  $M, \gamma, g$ in controlling the efficiency of the proposed single-photon quantum switch. The efficiency of the quantum switching, $\eta(t)$, is determined by the fraction of single-photon pulse leaked out to the TB waveguide within the time $t$. Thus, we define $\eta(t)=1-(|\psi_{e}(t)|^2+\sum \limits_{\substack{x=1}}^{M}\sum \limits_{\substack{\alpha=A,B}}|\psi_{x\alpha}(t)|^2)$. Since the tunneling between two edge modes is the most important ingredient for the emission leaking to the TB waveguide, $\eta(t)$ depends strongly on $M$. For a fixed set of $\gamma, g$ and topological phase of the SSH waveguide, there is an optimal $M$ for which $\eta(t)$ approaches unity in shorter times, as shown in Fig.~\ref{efficiency}(a). The appearance of an optimal $M$ (e.g., $M=9$ in Fig.~\ref{efficiency}(a)) in the topological phase can be explained through the following arguments. The hybridization of the edge modes is stronger for a shorter $M$, which generates a larger energy splitting of these modes. Then, the edge modes are more detuned from the emitted photon by the 2LE, resulting in lower $\eta(t)$. On the contrary, a longer $M$ reduces the hybridization of the edge modes as well as their energy detuning from the emitted photons. Simultaneously, a longer $M$ decouples the edge modes from each other, thus weakening the propagation channel. So $\eta(t)$ becomes smaller with longer $M$. Therefore, there is an optimal $M$ for simultaneous stronger coupling between the edge modes as well as between the 2LE and the edge modes. We further show the dependence of $\eta(t)$ on $g$ in Fig.~\ref{efficiency}(b) for fixed $M,\gamma$. 

The main features of $\eta(t)$ at short times in the topological phase of the SSH waveguide can be understood better by analytically studying the dynamics of a truncated model with the two edge modes of the SSH waveguide and the excited emitter. The truncated heterojunction model is developed in Appendix~\ref{AppendixD} by integrating all the modes of the TB waveguide and the bulk mode of the SSH waveguide. The integrating out of the TB waveguide in its semi-infinite limit ($N \to \infty$) generates a loss term (order of $-i\gamma^2/J$) at the edge mode near the interface. The loss term determines the emission of photons to the TB waveguide, and the emission probability increases with increasing $\gamma^2/J$. In the truncated model, we ignore the bulk modes of the SSH waveguide for a finite bulk gap when $v \ne w$. Further, the two edge modes of a finite SSH waveguide are tunnel coupled by amplitude $\xi~(\propto (v/w)^M)$ and the edge mode near the emitter is coupled to the emitter by amplitude $\zeta~(\propto g)$. These two coupling amplitudes set a timescale order of $1/(\sqrt{\xi^2+\zeta^2})$ for the oscillation of excitation between the edge modes of the SSH waveguide and the 2LE. In the truncated model, the timescale (order of $J/\gamma^2$) of photon emission to the TB waveguide through the interface competes with the timescale of excitation oscillation to settle the single-photon dynamics and the efficiency of the emission. When the emission timescale is shorter (e.g., for greater $\gamma$) than the oscillation timescale (e.g., for smaller $g$), the emission is favored, and $\eta(t)$ is higher. 

Fig.~\ref{efficiency}(b) indicates how an increasing $g$ reduces $\eta(t)$ by favoring the excitation oscillation between the SSH waveguide and the 2LE over leakage to the TB waveguide, which is controlled by $\gamma$. Fig.~\ref{efficiency}(b) also demonstrates that a higher $g$ increases the emission rate in the topologically trivial phase by coupling the 2LE to the TB waveguide through the bulk modes of the SSH waveguide as shown in Appendix~\ref{AppendixC}. 
In Appendix~\ref{AppendixC}, we apply Fermi's golden rule to derive the decay rate of an initially excited emitter into the TB waveguide by finding the wavefunction correction (up to 2nd order in $g$) to the emitter's subspace due to its finite coupling to the SSH waveguide in the trivial phase. We further compare the corresponding decay time scale with the time scale of the dynamics in the non-trivial phase given in Appendix ~\ref{AppendixD}.
 
\section{Summary and outlook}
 This paper carefully analyzes yet unexplored features of quantum light-matter interactions inside structured waveguides, mainly due to their finite bandwidths, band edges, and localized topological edge modes. We demonstrate that a well-known analogy between the transmission and reflection amplitude of a side-coupled 2LE to those of a direct-coupled 2LE for unstructured waveguides with infinite bandwidth no longer holds for the structured waveguide with finite bandwidth. This is due to the appearance of the Lamb shift for a 2LE direct coupled to waveguides with limited bandwidth. We further clarify the reasons for the emergence of the Lamb shift in different configurations of the emitter-waveguide coupling. 
It would be exciting to experimentally realize the direct-coupled 2LE for the TB and SSH waveguides and verify our theoretical predictions for the Lamb shifts and decay widths \cite{mirhosseini2018superconducting, du2022giant}. While nonreciprocity in multi-photon transport for asymmetric waveguide QED set-ups has been explored for unstructured waveguides \cite{PhysRevB.81.155117, FratiniPRL2014, RoyPRA2017, Hamann2018}, it is yet to be investigated for structured waveguides, especially for topological cases where the topological features can enhance the nonreciprocity. Our predictions for decoherence-free light-matter interactions with a giant 2LE near the band edges can readily be tested by combining various recent experimental advancements \cite{HoodKimble2016, LiuHouck2017, PhysRevX.11.011015, kannan2020waveguide}.   

Our investigation of single-photon dynamics in a heterojunction of two different structured waveguides using the first-principle numerics and the analytical method with a truncated model would be potentially valuable for future studies with large networks of these structured waveguides. While we have coupled an excited 2LE at the end of the heterojunction for the source of a photon, any other photon source can replace the 2LE in our study of single-photon dynamics. An extension of our research to multi-photon dynamics in such SSH waveguides and heterojunctions would be exciting for many physical phenomena, including the effect of topology on the correlated photon transport \cite{PhysRevLett.104.023602, PhysRevA.83.043823, PhysRevA.79.023837, ShenFanPRL2007, PhysRevB.81.155117, Zheng2010, Pletyukhov15, Manasi2018}, the resonance fluorescence \cite{Astafiev10a}, the Kerr and cross-Kerr effect \cite{HoiPRL2013, Vinu2020, Vinu2023}. 

\section{Acknowledgements}
We thank Ritu Nehra for helping us evaluate Green’s functions in Appendix~\ref{AppendixA}. 

\appendix
\section{Evaluation of Green's function}\label{AppendixA}
In Sec.~\ref{GiantSSH}, we have used the real space components of the retarded Green's function of the free Hamiltonian $\hat{H}_o=\hat{H}_e+\hat{H}_{SSH}$. We here explain how to find these components \cite{economou}.   
The completeness relation in the single excitation sector for the Hilbert space of $\hat{H}_o$ is
\begin{align}  \mathds{1}=|\varphi,e\rangle\langle \varphi,e|+\int_{-\pi}^{\pi}dq \big( |\phi_q^+\rangle\langle \phi_q^+|+|\phi_q^-\rangle\langle \phi_q^-| \big) |g\rangle \langle g|,\label{idenityresolution}
\end{align}
which we plug in the definition of the retarded Green's function in Eq.~\ref{LippmannRetardG} to obtain the following relation (the limit of $\epsilon\rightarrow0$ is assumed to be implicit in the following relevant relations): 
\begin{align}
   &G_o^{R}(E)=\frac{|\varphi,e\rangle\langle \varphi,e|}{\big(E-\omega_e+i\epsilon\big)}+\int_{-\pi}^{\pi}dq\sum_{x,x'} \frac{e^{iq(x-x')}}{4\pi} \nonumber\\ & \Bigg[\frac{\Big(e^{-i\theta_q}\,|x,A\rangle+|x,B\rangle\Big) |g\rangle \langle g| \Big(e^{i\theta_q}\,\langle x',A|+\langle x',B|\Big)}{\big(E-\omega_q^{+}+i\epsilon\big)}+\nonumber\\
   &\frac{\Big(-e^{-i\theta_q}\,|x,A\rangle+|x,B\rangle\Big) |g\rangle \langle g| \Big(-e^{i\theta_q}\langle x',A|+\langle x',B|\Big)}{\big(E-\omega_q^{-}+i\epsilon\big)}\Bigg].
\end{align}
The component of the retarded Green's function in the excited state of the 2LE is
\begin{align}
    \langle \varphi,e |G_o^{R}|\varphi,e\rangle= \frac{1}{\big(E-\omega_e+i\epsilon\big)}.
\end{align}
The components of the retarded Green's function in the real space basis of the SSH waveguide are:
\begin{align}
\langle g|\langle x,A |G_o^{R}|x',&B\rangle |g\rangle= 
\frac{1}{2\pi}\int_{-\pi}^{\pi}\frac{\big(v+we^{-i\,q}\big) e^{i q(x-x')}}{(E-\omega_o+i\epsilon)^{2}-f_q^{2}}dq, \label{GAB}\\
\langle g|\langle x,B |G_o^{R}|x',&A\rangle|g\rangle= \frac{1}{2\pi}\int_{-\pi}^{\pi}\frac{\big(v+we^{i\,q}\big) e^{i q(x-x')}}{(E-\omega_o+i\epsilon)^{2}-f_q^{2}} dq, \label{GBA}\\
\langle g|\langle x,A |G_o^{R}|x',&A\rangle|g\rangle=\langle g|\langle x,B |G_o^{R}|x',B\rangle|g\rangle \nonumber\\ =&\frac{1}{2\pi}\int_{-\pi}^{\pi}\frac{\big(E-\omega_o+i\epsilon\big) e^{i q(x-x')}}{(E-\omega_o+i\epsilon)^{2}-f_q^{2}} dq. \label{GAA}
\end{align}
The above integrals can be evaluated on the complex energy plane using the residue theorem. Let us define,
\begin{align}
I=&\frac{1}{2\pi}\int_{-\pi}^{\pi} \frac{e^{iq(x-x')}}{(E-\omega_o\;+i\epsilon)^{2}-f_q^{2}}dq \nonumber\\
=&\frac{1}{2\pi}\int_{-\pi}^{\pi} \frac{e^{iq(x-x')}}{(E-\omega_o)^{2}-f_q^{2}+2i\:(E-\omega_o)\epsilon} dq \nonumber\\
=&\frac{1}{2\pi\:vw}\int_{-\pi}^{\pi} \frac{e^{iq|x-x'|}}{2D-2\cos q+2i\:\frac{(E-\omega_o)}{vw}\epsilon}dq,\label{int}
\end{align}
where, $D=((E-\omega_o)^{2}-v^{2}-w^{2})/(2vw)$. We retain the terms up to first order in $\epsilon$ only. For $E$ inside the upper energy band of the SSH waveguides, we have $\epsilon' \equiv ((E-\omega_o)\epsilon)/(vw)\rightarrow0^{+}$ when $\epsilon\rightarrow0^{+}$ as we choose $vw>0$ and $E>\omega_o$. By substituting  $z=e^{iq}$, the integration in Eq.~\ref{int} can be cast into a contour integral over a unit circle around the origin of the complex plane.
\begin{align*}
I=&-\frac{1}{2\pi i\:vw}\oint \frac{z^{|x-x'|}}{-2(D+i\epsilon')z+z^{2}+1}dz\\
= &-\frac{1}{2\pi i\:vw}\oint \frac{z^{|x-x'|}}{(z-z_+)(z-z_-)}dz,
\end{align*}
where the integrand has two simple poles at
\begin{align}
    z_\pm =\big(D+i\epsilon'\big)\pm i\Big(\sqrt{1-D^{2}}-\frac{iD\epsilon'}{\sqrt{1-D^{2}}}\Big). \end{align} 
One of the poles, e.g., $z_{-}$ lies inside the unit circle. Thus, we find after applying the residue theorem and taking the limit $\epsilon\rightarrow0$:
\begin{align}
    I=-\frac{i}{2\,vw} \frac{(D-i\sqrt{1-D^{2}})^{|x-x'|}}{\sqrt{1-D^{2}}}.
\end{align}

Here, $\sqrt{1-D^{2}}$ is always positive. We substitute $D=\cos k$ in the above equation to derive
\begin{align}
    I=&\frac{-i}{2\:vw} \frac{(\cos k-i |\sin k |)^{|x-x'|}}{|\sin k|},
\end{align}
which is then used to find the components of the retarded Green's functions in Eqs.~\ref{GAB},\ref{GBA},\ref{GAA}. These are for $k<0$:
\begin{align}\label{SSHretardG}
 \langle g| \langle x,A |G_o^{R}|x',A\rangle |g \rangle=&\langle g|\langle x,B |G_o^{R}|x',B\rangle|g \rangle= \nonumber \\ &\frac{i}{2vw \sin{k}} |v+w e^{i k}| e^{i k|x-x'|},\\
\langle g|\langle x,A |G_o^{R}|x',B\rangle|g \rangle=& \frac{i}{2vw\sin{k}}\Big(ve^{ik|x-x'|}+w e^{i k|x-x'-1|}\Big), \\
\langle g|\langle x,B |G_o^{R}|x',A\rangle|g \rangle=& \frac{i}{2vw\sin{k}}\Big(v e^{i k|x-x'|} +w e^{i k|x-x'+1|}\Big).
\end{align}

\section{Self-energy correction of a 2LE direct coupled to the SSH waveguide} \label{AppendixB}
In the Sec.~\ref{decay} and Sec.~\ref{analogy}, we have identified the Lamb shift and the decay width of the emitter from the transport coefficients. In this appendix, we provide an alternative calculation for the Lamb shift and the decay width from the self-energy corrections of the 2LE due to its coupling to the waveguide modes. We particularly derive these for a 2LE direct coupled to the SSH waveguide in Sec.~\ref{dSSH}. The expressions for the Lamb shift and the decay width in Eqs.~\ref{transDSSH0},\ref{transDSSH} suggest that these appear as a sum of individual contributions from the left and right waveguide parts. Therefore, we can evaluate the self-energy correction from one side of  the waveguide, e.g., the left, and the other term due to the coupling to the right waveguide can be found by replacing $g_L$ by $g_R$ and swapping $(v,w) \rightarrow (w,v)$. 

Thus, we consider the Hamiltonian $\hat{H}_o+\hat{H}_g$ for the left SSH waveguide part, the emitter and their coupling by eliminating all the parts on the right of the 2LE from Eq.~\ref{dSSHHamiltonian}. We thus read from Eq.~\ref{dSSHHamiltonian}: $\hat{H}_{g}=g_L(\sigma^{+} b_{-1}+b_{-1}^{\dagger}\sigma)$. The bulk eigenfrequencies of the left SSH waveguide are the same as those Eq.~\ref{SSH dispersion}, and the corresponding eigenstates are obtained by fixing the boundary condition as $\langle 0,A|\phi_k^{\pm}\rangle_L=0$. 
\begin{align}
    |\phi_k^{\pm}\rangle_L = \sqrt{\frac{1}{\pi}}\sum_{x<0}\Big(\pm\sin{kx}|x,A\rangle+\sin{(kx+\theta_k)}|x,B\rangle\Big),
\end{align}
for $0<k<\pi$.
The self-energy correction $\Sigma_e$ to the 2LE is defined as \cite{cohen1998atom, steck2007quantum, PhysRevA.107.013710}
\begin{align}
\Sigma_e(E+i\,0^{+})=\lim_{\epsilon \rightarrow 0^{+}} \langle \varphi,e|\big(\hat{H}_{g}+\hat{H}_{g}\frac{\mathds{P}_{L}}{E-\hat{H}_o+i\epsilon}\hat{H}_{g}\big)|\varphi,e\rangle.
\end{align}
where the projection operator $\mathds{P}_{L}$ spans the single-excitation sector of the SSH waveguide modes and the ground state of the 2LE. $\mathds{P}_{L}=\int_{0}^{\pi}dk\big(|\phi_k^{+}\rangle_{LL}\langle \phi_k^{+}|+|\phi_k^{-}\rangle_{LL}\langle \phi_k^{-}|\big)|g\rangle \langle g|$ in the topologically trivial phase $(v>w)$ of the SSH waveguide, and  $\mathds{P}_{L}=\Big(\int_{0}^{\pi}dk\big(|\phi_k^{+}\rangle_{LL}\langle \phi_k^{+}|+|\phi_k^{-}\rangle_{LL}\langle \phi_k^{-}|\big)+|ed_1\rangle\langle ed_1|+|ed_2\rangle\langle ed_2|\Big)|g\rangle \langle g|$, where $|ed_{1}\rangle,|ed_{2}\rangle$ are two discrete localized modes with degenerate energy $\omega_{ed}=\omega_o$ in the topologically non-trivial phase when $v<w$. In the trivial phase without the edge states, we get for $\Sigma_{e}$:

\begin{align}  
&\Sigma_{e}(E+i\,0^{+})\nonumber\\
&=\lim_{\epsilon \rightarrow 0^{+}}\int_{0}^{\pi}dq\frac{\langle \varphi,e|\hat{H}_{g}|\phi_q^{+}\rangle_{L}|g\rangle \langle g|_L\langle \phi_q^{+}|\hat{H}_{g}|\varphi,e\rangle}{E-\omega_q^{+}+i \epsilon} \nonumber\\
&\quad+\lim_{\epsilon \rightarrow 0^{+}}\int_{0}^{\pi}dq\frac{\langle \varphi,e|\hat{H}_g|\phi_q^{-}\rangle_{L}|g\rangle \langle g|_L\langle \phi_q^{-}|\hat{H}_g|\varphi,e\rangle}{E-\omega_q^{-}+i \epsilon}\nonumber\\
 &=\frac{1}{\pi}\lim_{\epsilon \rightarrow 0^{+}}\int_{0}^{\pi}dq\frac{1}{E-\omega_q^{+}+i\epsilon}\frac{g_L^2v^2\sin^2{q}}{f_q^2} \nonumber\\
 &\quad\quad +\frac{1}{\pi}\lim_{\epsilon \rightarrow 0^{+}}\int_{0}^{\pi}dq\frac{1}{E-\omega_q^{-}+i \epsilon}\frac{g_L^2v^2\sin^2{q}}{f_q^2} \nonumber\\
&=\frac{1}{2\pi}\lim_{\epsilon \rightarrow 0^{+}}(E-\omega_o+i\epsilon)\int_{-\pi}^{\pi}dq\frac{g_L^2v^2(1-\cos{2q})}{(E-\omega_o+i\epsilon)^{2}-f_q^{2}}\frac{1}{f_q^2}\nonumber\\
 &=\frac{1}{2\pi}\lim_{\epsilon \rightarrow 0^{+}}(E-\omega_o+i\epsilon)\int_{-\pi}^{\pi}dq\frac{g_L^2v^2(1-e^{i2q})}{(E-\omega_o+i\epsilon)^{2}-f_q^{2}}\frac{1}{f_q^2}. \label{SigT} 
\end{align}
The above integration can be cast into an integral on the complex plane by substituting $z=e^{iq}$. After evaluating the integration in Eq.~\ref{SigT} following the procedure in Appendix~\ref{AppendixA}, we find: 
\begin{align}
&\Sigma_{e}(E+i\,0^{+}) \label{B4}\\
&=\begin{cases}
       \frac{i}{2vw\,\sin{k}}\frac{1}{f_k}g_L^2v^2\big(1-e^{i2k}\big) + \frac{g_L^2}{f_k},~~~~~ \text{ for } v>w, \\
       \frac{i}{2vw\,\sin{k}}\frac{1}{f_k}g_L^2v^2\big(1-e^{i2k}\big) + \frac{v^2}{w^2}\frac{g_L^2}{f_k},~ \text{ for } v<w.
\end{cases} \nonumber 
\end{align}

We here note that we have parameterized above $E=\omega_o+f(k)$, with $k<0$. 
We further need to include in $\Sigma_{e}(E+i\,0^{+})$ the contribution from the edge modes when $v<w$. One of these two modes (say $|ed_1\rangle$) has a non-zero wavefunction amplitude at the SSH waveguide boundary site $(x,\alpha)=(-1, B)$, where the 2LE is coupled (the other is located at the other boundary of unit cell $x=-\infty$). The wavefunction of the edge mode $|ed_1\rangle$ in real space basis is given by,
\begin{align}
    |ed_1\rangle=\sqrt{1-\frac{v^2}{w^2}}\sum_{x \geq 1} \Big(-\frac{v}{w}\Big)^{x-1}|-x,B\rangle.
\end{align}
The self-energy correction $\Sigma_{e}^{ed}(E+i 0^{+})$ due to the topological edge modes in the non-trivial phase is
\begin{align}
&\Sigma_{e}^{ed}(E+i 0^{+})\nonumber\\
&=\frac{|\langle \varphi,e|\hat{H}_{g}|ed_1 \rangle |g\rangle|^2}{E-\omega_{ed}+i 0^{+}}
=\frac{g_L^2}{f_k}\Big(1-\frac{v^2}{w^2}\Big).\label{B7}
\end{align}
Adding the contributions in Eq.~\ref{B4} and Eq.~\ref{B7} for $v<w$, we find that the total self-energy correction $\Sigma_{e}$ is the same in both the topological phases, and it is given by
\begin{align}
    \Sigma_{e}(E+i\,0^{+})&=\frac{i}{2\,vw\sin{k}}\frac{1}{f_k}\;g_L^2\,v^2\;\big(1-e^{i\,2k}\big)+\frac{g_L^2}{f_k}\nonumber\\
    &= \frac{g_L^2}{f_k}\frac{v}{w}\big( \cos{k}+i\sin{k}\big)+\frac{g_L^2}{f_k}.
\end{align}
We can identify the Lamb shift ($\tilde{\Delta}_L$) and the decay width ($\tilde{\Gamma}_L$) from $\Sigma_{e}(E+i\,0^{+})$ expressed as $\Sigma_{e}(E+i\,0^{+})=\tilde{\Delta}_L-i\tilde{\Gamma}_L/2$: 
\begin{align}
    \tilde{\Delta}_L&=\frac{g_L^2}{w\,f_k}\big(w+v\cos{k}\big),\\
\frac{\tilde{\Gamma}_L}{2}&=-\frac{g_L^2}{f_k}\frac{v}{w}\sin{k}.
\end{align}

\section{Truncated model for single-photon dynamics in the topological phase of the heterojunction coupled to 2LE}\label{AppendixD}
Here we explain how we obtain the truncated model discussed in Sec.~\ref{Qswitch} for the single-photon dynamics in the topological phase of the heterojunction coupled to 2LE.  We integrate out all the modes of the semi-infinite TB waveguide and the bulk mode of the SSH waveguide in two steps. First, we integrate out of the TB modes as we describe below. An eigenstate of the full Hamiltonian of the system $\hat{H}_T$ in the limit of $N\rightarrow \infty$ in the single-excitation sector can be written as,
\begin{align}
|\tilde{\psi}\rangle=\int_{0}^{\pi} dk \tilde{\psi}_k |k\rangle |g\rangle +\sum_{x=1}^{M}&(\tilde{\psi}_{xA}|x,A\rangle
+\tilde{\psi}_{xB}|x,B\rangle)|g\rangle\nonumber\\+&\tilde{\psi}_e|\varphi,e\rangle,
\end{align}
where, we employ the real-space and the momentum-space basis states for the SSH and the TB waveguide, respectively. Here, $\tilde{\psi}_k,\tilde{\psi}_{xA}$ and $\tilde{\psi}_{xB}$ are the respective amplitude of a single photon in the $k^{\rm th}$ mode of the TB waveguide, at $A$ and $B$ sublattice of $x^{\rm th}$ unit cell of the SSH waveguide when the 2LE in the ground state. The momentum-space basis states of the TB waveguide (for $N \to \infty$) are:
\begin{align}
|k\rangle=\sqrt{\frac{2}{\pi}}\sum_{x\leq 0} \sin{k(x-1)}\, |x\rangle. \label{c2}
\end{align} 
We find the following relations by taking projection of $\hat{H}_T|\tilde{\psi}\rangle=E|\tilde{\psi}\rangle$ on the TB modes $|q\rangle|g\rangle$ and the SSH sublattice $|1,A\rangle |g\rangle$:
\begin{align}
    &\tilde{\psi}_q=-\gamma \sqrt{\frac{2}{\pi}}\frac{\sin{q}}{E-\omega_q}\tilde{\psi}_{1A}, \label{psiq}\\
    &v\tilde{\psi}_{1B}-\gamma\int_{0}^{\pi} dq \sqrt{\frac{2}{\pi}}\sin{q}\,\tilde{\psi}_q=(E-\omega_o)\tilde{\psi}_{1A}.\label{psia}
    \end{align}
We substitute $\tilde{\psi}_q$ from Eq.~\ref{psiq} in Eq.~\ref{psia} and get:
   \begin{align}
  v\tilde{\psi}_{1B}=\Big(E-\omega_o-\gamma^2\frac{2}{\pi}\int_{0}^{\pi} dq\frac{\sin^2{q}}{E-\omega_q}\Big)\tilde{\psi}_{1A}.\label{D2}
\end{align}
We evaluate the integration in Eq.~\ref{D2} by shifting the pole of the integrand into the complex plane, and transforming the integration into a complex integral. We thus find:
\begin{align}
   \lim_{\epsilon\rightarrow 0^+} \frac{2}{\pi}\int_{0}^{\pi} dq\frac{\sin^2{q}}{E-\omega_q+i\epsilon}=\frac{1}{J}\Big(\cos{k}-i\sin{k}\Big),
\end{align}
where, $E=\omega_o+2J\cos{k} \equiv \omega_k$, with $k>0$. Therefore, we get from Eq.~\ref{D2}:
\begin{align}
    \tilde{\psi}_{1A}=\frac{v\tilde{\psi}_{1B}}{\omega_k-\big(\omega_o+\Delta_k-i\Gamma_k/2\big)},\label{D5}
\end{align}
where, $\Delta_k=\gamma^2\cos{k}/J$  and $\Gamma_k/2 =\gamma^2\sin{k}/J$ are the real and imaginary frequency shifts, respectively, to the end sublattice site $(1,A)$ of the SSH waveguide due to the integration out of the TB band continuum. For $\gamma^2/J \ll \omega_o$, we approximate $\omega_k \simeq \omega_o$, which gives $k \approx \pi/2$ and $\Delta_{k \approx \pi/2}=0$ and $\Gamma_{k \approx \pi/2}/2 =\gamma^2/J$. Within this approximation, the end $(1,A)$ sublattice's energy acquires only an imaginary shift or a finite decay width. 
Therefore, we can write the following effective Hamiltonian for the full system by integrating the TB waveguide out: 
\begin{align}
\hat{H}_{1}=\hat{H}_{SSH}+\hat{H}_{e}+\hat{H}_{I}-i\big(\frac{\gamma^2}{J}\big) a^{\dagger}_1a_1.
\end{align}
Next, we eliminate the bulk modes of the SSH waveguide in the topological phase when $g^2/J$ is small in comparison with the frequency differences of the 2LE and the SSH bulk modes for a finite bulk gap (e.g., $v<w$). We are essentially thus left with three modes of the full system: $|e\rangle$ of the 2LE and the two edge modes ($|ed^{\pm}\rangle$) \cite{PhysRevB.84.195452} of the SSH waveguide. Further, we choose to work with the symmetric (anti-symmetric) linear combination of the edge modes, $|L\rangle=(|ed^+\rangle +  |ed^-\rangle)/\sqrt{2}~(|R\rangle=(|ed^+\rangle-|ed^-\rangle)/\sqrt{2})$, which ensure localized state on the sublattice $A~(B)$ at the left (right) boundary of the SSH waveguide. The final effective Hamiltonian of the truncated model in the basis of $\{|L\rangle |g \rangle, |R\rangle|g\rangle,|\varphi,e\rangle\}$ is 
\begin{align}
\hat{H}_{\rm eff}=\begin{pmatrix}
    \omega_o-i\Upsilon & \xi &0\\
    \xi & \omega_o & \zeta\\
    0& \zeta & \omega_o
\end{pmatrix},\label{Non-hermitian}
\end{align}
where, $\Upsilon=\gamma^2 |\langle L|1,A\rangle |^2/J$ determines the decay rate of excitation from the left edge state, $\xi= \langle L|\hat{H}_{SSH}|R\rangle $ denotes the overlap of the left and right edge states, and $\zeta=g\langle R|M, B\rangle $ is the effective coupling of the 2LE with the right edge state of the SSH waveguide. We here follow the treatment in \cite{asboth2016short} to approximately estimate $\Upsilon,\zeta$ and $\xi$ as given below,
\begin{align}
    \Upsilon \approx \frac{\gamma^2}{J}&\frac{1-\frac{v^2}{w^2}}{1-\big(\frac{v^2}{w^2}\big)^M},\quad \zeta \approx g \sqrt{\frac{1-\frac{v^2}{w^2}}{1-\big(\frac{v^2}{w^2}\big)^M}}, \nonumber\\ &\xi\approx w \big(\frac{v}{w}\big)^M \frac{1-\frac{v^2}{w^2}}{1-\big(\frac{v^2}{w^2}\big)^M}.
\end{align}
By treating the non-Hermitian term in Eq.~\ref{Non-hermitian} as a perturbation for $\Upsilon < \sqrt{\xi^2+\zeta^2}$, we find the right eigenvalues of $\hat{H}_{\rm eff}$ upto first order correction in $\Upsilon$ as 
\begin{align}
    &\omega_0\approx \omega_o-i\frac{\zeta^2}{\xi^2+\zeta^2}\Upsilon, \\
    &\omega_{\pm} \approx \omega_o \pm \sqrt{\xi^2+\zeta^2}-i\frac{\xi^2}{2(\xi^2+\zeta^2)}\Upsilon.
\end{align}
\begin{figure}
\centering
\includegraphics[width=0.47\textwidth]{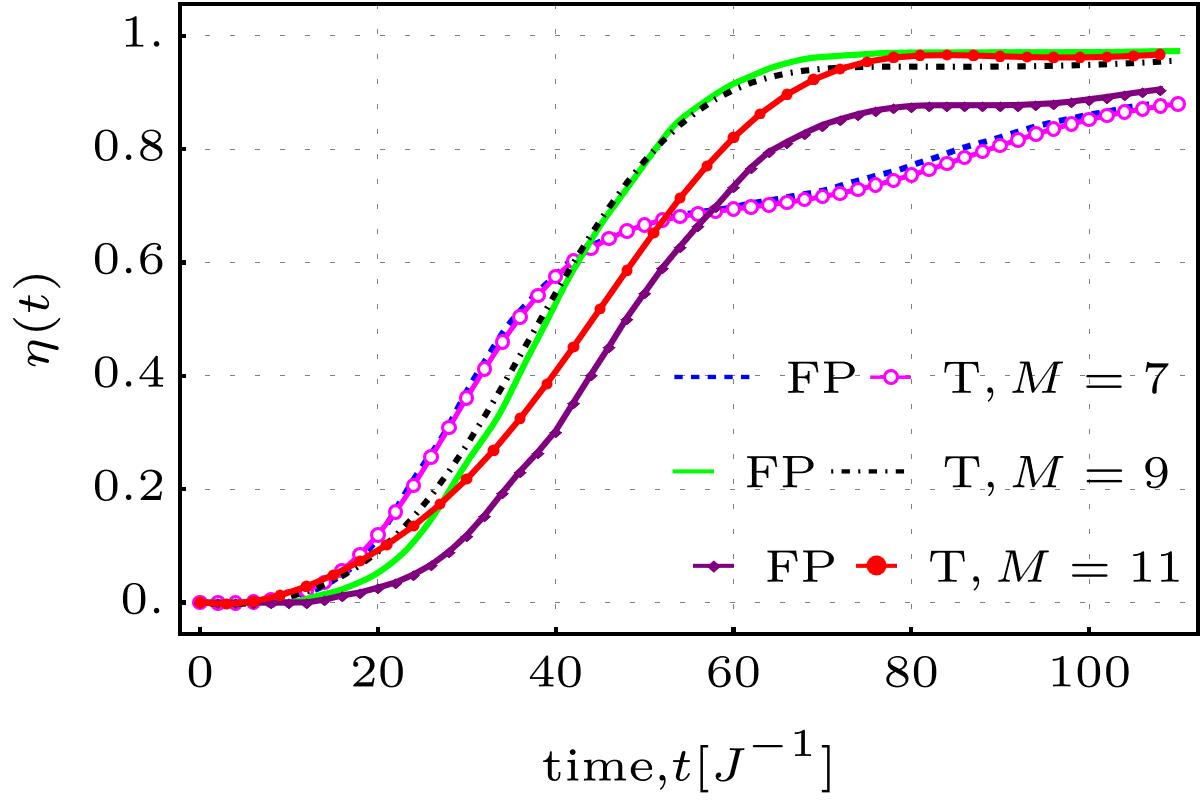}
\caption{Comparison of photon emission efficiency $\eta(t)$ with $t$ calculated from the first-principle numerics (FP) in Sec.~\ref{Qswitch} and the truncated model (T) in Sec.~\ref{AppendixD}. The parameters are, $g/J=0.07$, $\gamma/J=0.4$, $\omega_o/J=3.0$, $t_{\rm f}=110/J$, $\delta t=0.001/J$, $N=119$ and $\delta=0.1$ indicating the topological phase of the SSH waveguide. }\label{Analy non-Hermitian}
\end{figure}
The corresponding eigenvectors are $|\omega_{j} \rangle \approx (A_{j},B_j,C_j)/\sqrt{2(\xi^2+\zeta^2)}$ for $j=0,\pm$, with,
\begin{align}
    &A_0=-\sqrt{2}\zeta,B_0=-i\frac{\sqrt{2} \zeta \xi }{\xi^2+\zeta^2}\Upsilon,C_0=\sqrt{2}\xi,\\
   &A_{\pm}=\xi\Big(1\mp i\frac{\xi^2+4\zeta^2}{4(\xi^2+\zeta^2)^{\frac{3}{2}}}\Upsilon \Big),\\
   &B_{\pm}=\pm \sqrt{\xi^2+\zeta^2}+i\frac{\xi^2}{4(\xi^2+\zeta^2)}\Upsilon,\\
    &C_{\pm}= \zeta \Big(1\pm i\frac{3\xi^2}{4(\xi^2+\zeta^2)^{\frac{3}{2}}}\Upsilon\Big).
\end{align}
Applying the truncated model, we finally calculate the wavefunction for single-photon dynamics in the topological phase for an excited emitter as an initial condition. The time evolution of the wavefunction is given $|\psi_T(t)\rangle=\sum\limits_{\substack{j=0,\pm}} y_j e^{-i\omega_j^{}t} |\omega_{j}^{}\rangle$, where
 \begin{flalign}
    &y_0=\frac{\sqrt{2(\xi^2+\zeta^2)} B_{-} (B_{+}A_{-}-A_{+}B_{-})}{\splitfrac{(A_0B_{-}-B_0A_{-})(B_{-}C_{+}-C_{-}B_{+})}{-(C_{-}B_0-C_0B_{-})(B_{+}A_{-}-B_{-}A_{+})}},\\
     &y_{+}=\frac{A_0B_{-}-B_0A_{-}}{B_{+}A_{-}-A_{+}B_{-}}y_0,y_{-}=-\frac{A_0y_0+A_{+}y_{+}}{A_{-}}.
 \end{flalign}
 The photon emission efficiency is now given by $\eta(t)=1-|\psi_T(t)|^2$, which is compared in Fig.~\ref{Analy non-Hermitian} with that obtained from the first-principle numerics in Sec.~\ref{Qswitch}.  Fig.~\ref{Analy non-Hermitian} shows a good agreement between two different calculations of $\eta(t)$ in the topological phase for small $g$ when the bulk modes of the SSH waveguide do not contribute in the single-photon dynamics.

\section{Decay rate of an excited 2LE coupled to the heterojunction in the trivial phase of SSH waveguide}\label{AppendixC}
\begin{figure}
\centering
\includegraphics[width=0.47\textwidth]{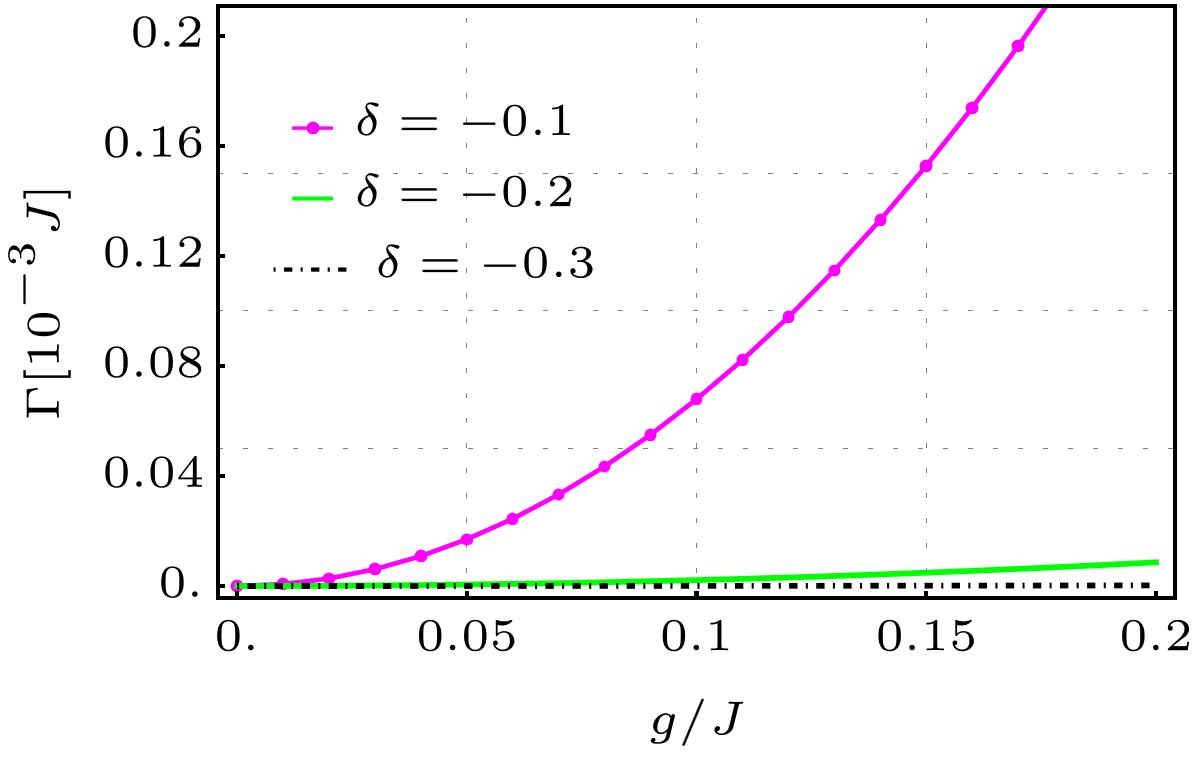}
\caption{Variation of decay rate $\Gamma$ with coupling amplitude $g$ for different $\delta$ in the trivial phase of the SSH waveguide. The decay rate decreases as the $\delta$ increases for a given $g$. Other parameters are $\gamma/J=0.4$, $M=9$.}\label{decay in trivial phase}
\end{figure}

In Fig.~\ref{efficiency}(b) of Sec.~\ref{Qswitch}, we have shown how the efficiency, $\eta(t)$, of the quantum switch, varies with the coupling amplitude $g$ in different topological phases of the SSH waveguide. We observe the photon emission rate improves as $g$ increases in the topologically trivial phase.  Applying Fermi's golden rule, we here derive an expression for the decay rate  $\Gamma$ of an excited 2LE into the quasi-continuum of the  TB waveguide for the SSH waveguide in its trivial phase. The decay rate is 
\begin{align}
    \Gamma=2\pi |\langle g| \langle k|\hat{H}_{\gamma}|\tilde{e}\rangle|^2\rho(\omega_k)|_{\omega_k=\omega_e}.\label{Fermi golden}
\end{align}  
Here, $|\tilde{e}\rangle$ is the perturbed wave-function of the excited emitter due to its coupling to the SSH waveguide. The state $|k\rangle$ is given in Appendix~\ref{AppendixD}.  
We fix $\omega_e=\omega_o$ in our problem. Thus the local density of states around $\omega_k=\omega_o$ at site $x=0$ of the TB waveguide is $\rho(\omega_k)= \sin{k} / (\pi J )=(\pi J)^{-1}$. An expression for $|\tilde{e}\rangle$ is obtained by including the perturbative correction up to the 2nd order of the coupling amplitude $g$ of the 2LE with the SSH waveguide:
\begin{align}
   |\tilde{e}\rangle =  |\varphi,e\rangle&+\sum_{q} \Big\{ \frac{|\phi_q^{+}\rangle |g\rangle \langle g |\langle \phi_q^{+}|\hat{H}_I|\varphi,e\rangle}{\omega_e-\omega_q^{+}} \nonumber\\
   &+\frac{|\phi_q^{-} \rangle|g\rangle \langle g |\langle \phi_q^{-}|\hat{H}_I|\varphi,e\rangle}{\omega_e-\omega_q^{-}} \Big\} +O(g^3),\label{C3}
\end{align}
where  $q$ takes $M$ values for the bulk modes of the SSH waveguide in the trivial phase. These bulk modes for a finite SSH chain are given in \cite{PhysRevB.84.195452},
\begin{align} &|\phi_q^{\pm}\rangle=c_q\sum_{x=1}^{M}\Big(\pm\sin{(xq-\theta_q)}|x,A\rangle +\sin{xq}|x,B\rangle\Big),\label{C4}
\end{align}
where $c_q$'s are normalization constants. The values of $q$'s are obtained from the boundary condition $\langle M+1,A|\phi_k^{\pm}\rangle=0$, that gives the following transcendental equation, $v\sin{(M+1)q}=-w\sin{Mq}$, which can be solved numerically. Plugging Eqs.~\ref{c2},\ref{C3} and \ref{C4} in Eq. \ref{Fermi golden}, the decay rate in the trivial phase of the SSH waveguide is obtained as 
 \begin{align}
\Gamma = \frac{16\gamma^2 g^2}{\pi J}\Big|\sum_q c_q^2&\frac{\sin{Mq}\sin{(q-\theta_q)}}{f(q)} 
 \Big|^2.
 \end{align}
The above expression shows $\Gamma \propto g^2$, which indicates an increment of the single photon emission to the TB waveguide for a higher $g$ due to stronger emitter coupling with the TB waveguide through the bulk modes of the finite SSH waveguide in the trivial phase. Fig ~\ref{decay in trivial phase} displays the quadratic dependence of $\Gamma$ on $g$ for different values of $\delta$. Since the bulk modes are further detuned from the 2LE transition frequency $\omega_e$ when $\delta$ is increased, the decay rate is reduced drastically. Next, we compare the timescale ($1/\Gamma$) associated with the decay of an excited 2LE in this trivial phase to the timescale in the non-trivial phase via the edge modes as discussed in Appendix ~\ref{AppendixD}. For $\delta=-0.1$,  $\gamma/J=0.4$, $g/J=0.07$, and $M=9$, we find $1/\Gamma \approx 3 \times 10^{4} J^{-1}$, which is nearly $690$ times longer than  the timescale of emission in Fig ~\ref{efficiency}.  

\bibliography{references} 

\end{document}